 \documentclass[]{revtex4-2}

\usepackage{lipsum}
\makeatletter
\def\ps@pprintTitle{%
 \let\@oddhead\@empty
 \let\@evenhead\@empty
 \def\@oddfoot{}%
 \let\@evenfoot\@oddfoot}
\makeatother

\usepackage{subcaption}
\usepackage[utf8]{inputenc}
\usepackage[export]{adjustbox}
\usepackage{wrapfig}
\usepackage{graphicx}
\usepackage{amssymb}
\usepackage{amsmath, nccmath}
\usepackage{commath}
\usepackage{amsthm}
\usepackage{geometry}
\usepackage{lineno}
\makeatletter
\newcommand*{\rom}[1]{\expandafter\@slowromancap\romannumeral #1@}

\usepackage{lipsum}
\usepackage{array}
\makeatletter
\def\ps@pprintTitle{%
 \let\@oddhead\@empty
 \let\@evenhead\@empty
 \def\@oddfoot{}%
 \let\@evenfoot\@oddfoot}
\makeatother

\begin{document}



\title{Continued functions and critical exponents: Tools for analytical continuation of divergent expressions in phase transition studies}



\author{Venkat Abhignan, R.  Sankaranarayanan}
\address{Department of Physics, National Institute of Technology, Tiruchirappalli-620015, India. \\
yvabhignan@gmail.com, sankar@nitt.edu.}
\begin{abstract}
Resummation methods using continued functions are implemented to converge divergent series appearing in perturbation problems related to continuous phase transitions in field theories. In some cases, better convergence properties are obtained using continued functions than diagonal Pad\'e approximants, which are extensively used in literature. We check the reliability of critical exponent estimates derived previously in universality classes of $O(n)$-symmetric models (classical phase transitions) and Gross-Neveu-Yukawa models (quantum phase transitions) using new methods.
\end{abstract}





\maketitle
\section{Introduction}
Divergent series are inevitable solutions of perturbation approximations used in field theories \cite{Dyson}. Resummation methods are required to extract meaningful values from these perturbative expansions with zero radii of convergence around their singular points \cite{CALICETI2007,kleinert}. A summation method expands this region of convergence by following a different mapping of variables. The rigorous analysis by Stieltjes on continued fractions has led to the applicability of its analogue Pad\'e sequences on a wide range of problems in perturbation theory \cite{bender1999advanced}. Pad\'e based methods are the most commonly used techniques to achieve these affine transformations of variables by scaling and shifting \cite{Baker,baker_graves-morris_1996}. Our previous work showed that even other continued functions had remarkably interesting convergence properties by obtaining results related to universal critical parameters \cite{abhignan2020continued}. Some important results were discussed, implementing only the lower order information of the renormalization group (RG) perturbative expansions in the $O(n)$-symmetric $\phi^4$ scalar field theory. Especially using the continued exponential \cite{contexp} and such a blended function, continued exponential fraction, we could address the $\lambda$-point discrepancy between the theoretical predictions \cite{MC2006,Chester2020,nprg,mcn2}, and famous experimental value of specific heat exponent \cite{LIpa2003} in the $O(2)$ $\phi^4$ model \cite{lambda,shalaby2020critical}, though the issue remains unresolved. 

Also, using the continued exponential fraction, a consensus can be seen between different theoretical approaches in the most prominently solved three-dimensional Ising model where correlation length exponent $\nu_{Ising}\approx0.630$ matches up to the third decimal place. The different significant approaches in such models are perturbative RG \cite{sevenloop}, Monte-Carlo simulations (MC) \cite{MCHAS}, and conformal bootstrap calculations (CB)\cite{Kos2016}. Further, using these continued functions and combining them with Borel-Leroy transformation, we could produce precise estimates for critical parameters in universality classes of modified Landau-Wilson Hamiltonian \cite{abhignan2021}. Perturbative six-loop $\epsilon$ expansions from $n$-vector model with cubic anisotropy \cite{ADZHEMYAN2019}, $O(n)\times O(m)$ spin models \cite{KOMPANIETS2020} and the weakly disordered Ising model \cite{RIM2021} were handled. 

The simplest description for a sequence of the continued functions where the convergent behaviour is observed is that $(i+1)$th term of the sequence has the form of $i$ iterations of a corresponding function. Using the self-similar continued representation of a function to obtain convergence was the rudimentary idea developed into many forms by Yukalov \cite{Yukalov2019,physics3040053}. Even the recently developed resummation methods to achieve analytic continuation are based on orthogonal Gauss hypergeometric functions \cite{hgm1,hgm2,hgm3,shalaby2020critical}, which can be represented in the form of continued fractions \cite{Vleck}. However, such methods are based on using the large-order behaviour of the perturbative expansions. Since this asymptotic information might not be available in all cases, it is of prime importance to study resummation methods that only implement lower-order information. For instance, with the recent development in computational techniques, such lower-order information for the $\phi^4$ field theory has been solved to calculate the six-loop \cite{sixloop} and seven-loop \cite{sevenloop} RG functions. Such calculations involve around 138 Feynman graphs in the fifth order, 687 graphs in the sixth order and 4047 graphs in the seventh order in the perturbative expansions using the minimal subtraction renormalization scheme \cite{sixloop}. With the possibility of solving such complex calculations, it can lead to more orders of information in such field theories, which can better define the behaviour of classical and quantum phase transitions on a wide range of physical systems. 

Initially in Sec. \rom{2}.A we briefly introduce the description of resummation methods. We elaborated and implemented the resummation procedures using continued functions to evaluate divergent expressions concerning the correction-to-scaling exponents derived from the $\phi^4$ field theory. Previously we used the continued fraction to solve this series, which can be related to the diagonal Pad\'e approximant in orthogonal polynomials \cite{Aptekarev_2011,Bultheel2001,LORENTZEN20101364}. Further in Sec. \rom{2}.B, we have explored the role of continued exponential in other aspects of continuous phase transitions related to the lattice Ising model. Studying continuous phase transitions on a one-dimensional lattice model with short-range interaction was first proposed by Ising \cite{1925ZPhy...31..253I}. We implement the continued exponential into the schemes of widely studied perturbative low-temperature expansion \cite{domb1960} and primitive position-space renormalization \cite{Kadanoff:1976jb} of the Ising model. Finally, in Sec.\rom{3}, we explore the possibility of implementing continued functions in the study of critical exponents related to quantum phase transitions using RG functions of Gross-Neveu-Yukawa models \cite{4loopgny}.

\section{Critical exponents of Classical phase transitions \\ ($O(n)$ universality class)}
\subsection{$O(n)$-symmetric $\phi^4$ field theory and resummation of critical exponents}
Studying continuous phase transitions through $\phi^4$ field theory begins from Landau's description \cite{Landau:1937obd}. The most interesting numerical results that can be derived from implementing Kadanoff's scaling theory \cite{KADANOFF:1967zz}, Wilson's perturbative RG, and epsilon expansion \cite{Wilson:1973jj} to this theoretical description of Landau are the critical exponents. They describe the singular behaviour of the phase transition in a material at the critical point $T=T_c$ (critical temperature) and are considered universal, i.e., they are independent of the nature of the material and the type of continuous phase transition. These universal critical parameters are dependent only on the symmetries of the system and dimensionality $d$, defined by a universality class. These physically relevant yet numerically divergent critical exponents are solved from field theories in the form of \begin{equation}
    Q(\epsilon)\approx\sum^N_{i=0} q_i \epsilon^i
\end{equation} for $\epsilon\rightarrow0$ where $\epsilon$ is the perturbative parameter associated with the physical system.\\ Transformations of sequences is a key numerical technique for resolving convergence issues in the divergent series of critical exponents. The idea behind resummation techniques is that one can achieve convergence by combining the infinite divergent series with an appropriate sequence transformation rather than simply adding a particular series term by term, which is meaningless \cite{CALICETI2007}. A slowly converging or diverging sequence $\{s_N\}_{N=0}^{\infty}$, with the partial sums $\{s_N\}=\sum_{i=0}^N q_i$ of an infinite series, is transformed into a new, presumably better numerically behaved sequence, $\{s'_N\}_{N=0}^{\infty}$ using these resummation methods. Assume that $\{s_N\}_{N=0}^{\infty}$ either converges to a limit $s$ or, if it diverges, can be resummed using the right technique to produce $s$. Resummation methods implement transformations of linear sequences according to the general formula $s'_N=\sum_{i=0}^N \ \mu_{Ni} \ s_i$. These transformations compute the elements of the transformed sequence $\{s'_N\}$ as weighted averages of the elements of the input sequence $\{s_N\}$ with weights $\mu_{Ni}$. The primary argument is that, for the weights $\mu_{Ni}$, it is possible to establish some necessary and sufficient conditions to ensure that, when applied to a convergent sequence $\{s_N\}_{N=0}^{\infty}$, the converted sequence $\{s'_N\}_{N=0}^{\infty}$ may converge to the same limit, $s = s_{\infty}$. Depending on which resummation method is chosen for a specific problem, some empirical ideas are always required to get the best results. \\
We implement transformation of sequences using  continued exponential fraction \cite{abhignan2020continued, abhignan2021} \begin{equation}
   Q(\epsilon) \sim
 c_0\exp\left(\frac{1}{1+c_1\epsilon\exp\left(\frac{1}{1+c_2\epsilon\exp\left(\frac{1}{1+c_3\epsilon\exp\left(\frac{1}{1+\cdots}\right)}\right)}\right)}\right),  \end{equation} continued exponential \begin{equation}
    Q(\epsilon) \sim d_0\exp(d_1\epsilon \exp(d_2 \epsilon \exp(d_3 \epsilon \exp(d_4 \epsilon\exp(d_5 \epsilon\exp(d_6 \epsilon\exp(\cdots))))))), \end{equation} continued exponential with Borel-Leroy transformation \cite{abhignan2021} \begin{equation}
        Q(\epsilon) \sim \int_0^\infty \exp(-t) t^l e_0\exp(e_1\epsilon t\exp(e_2\epsilon t\exp(e_3\epsilon t\exp(\cdots)))) dt
    \end{equation} where $l$ is Borel-Leroy parameter, continued natural logarithmic function \begin{equation}
Q(\epsilon) \sim \log\left(g_1\epsilon\,\log\left(g_2\epsilon\log\left(g_3\epsilon\log\left(g_4\epsilon\log\left(g_5\epsilon\log\left(g_6\epsilon\log\left(\cdots\right)+1\right)+1\right)+1\right)+1\right)+1\right)+1\right),
    \end{equation} and continued fraction \begin{equation}
    Q(\epsilon) \sim \frac{h_0}{\frac{h_1\epsilon}{\frac{h_2\epsilon}{\frac{h_3\epsilon}{\frac{h_4 \epsilon}{\cdots}+1}+1}+1}+1},  \; \; \; \hbox{for} \; \; \; (\epsilon \rightarrow 0).
\end{equation} As was previously indicated, Yukalov frequently utilised such continued functions and their combinations for studying convergence \cite{Yukalov2019,physics3040053}. Also, as mentioned earlier, a continued fraction is related to the diagonal Pad\'e approximants and is easily manipulable algebraically \cite{Aptekarev_2011,Bultheel2001,LORENTZEN20101364}. Bender and Vinson were the first to investigate continued exponential \cite{contexp}, and it was later employed for studying convergence in phase transitions \cite{POLAND1998394,abhignan2020continued,abhignan2021}. Combining continued exponential and continued fraction, continued exponential fraction was utilised \cite{abhignan2020continued}. Continued exponential with Borel-Leroy transformation was utilised based on the Pad\'e-Borel-Leroy transformation \begin{equation}
     Q(\epsilon) = \int_0^\infty \exp(-t) t^l f(\epsilon t) dt ,\,\,\, f(y) = \sum_{i=0}^\infty \frac{q_i}{\Gamma(i+l+1)} y^i,
 \end{equation} replacing Pad\'e with continued exponential \cite{abhignan2021}. Factorial growth of coefficients $q_i$ can be determined as $i^l i!$ similar to $\Gamma(i+l+1)$ in the above equation using Stirling's approximation for large order behaviour ($i \rightarrow \infty$) \cite{kleinert}. \\ The convergence is noted by numerically observing the transformed sequence of calculated quantities \begin{multline}
     C_1 \equiv c_0\exp\left(\frac{1}{1+c_1\epsilon}\right),\,C_2 \equiv c_0\exp\left(\frac{1}{1+c_1\epsilon\exp\left(\frac{1}{1+c_2\epsilon}\right)}\right),\\ C_3 \equiv c_0\exp\left(\frac{1}{1+c_1\epsilon\exp\left(\frac{1}{1+c_2\epsilon\exp\left(\frac{1}{1+c_3\epsilon}\right)}\right)}\right),\cdots
 \end{multline} for continued exponential fraction,
 \begin{equation}
    D_1 \equiv d_0\exp(d_1\epsilon),\,D_2 \equiv d_0\exp(d_1\epsilon\exp(d_2\epsilon)),\,D_3 \equiv d_0\exp(d_1\epsilon\exp(d_2\epsilon\exp(d_3\epsilon))),\,\cdots
\end{equation} for continued exponential, \begin{multline}
     E_1 \equiv \int_0^\infty \exp(-t) t^l e_0\exp(e_1\epsilon t) dt,\,E_2 \equiv \int_0^\infty \exp(-t) t^l e_0\exp(e_1\epsilon t\exp(e_2\epsilon t)) dt,\\ E_3 \equiv \int_0^\infty \exp(-t) t^l e_0\exp(e_1\epsilon t\exp(e_2\epsilon t\exp(e_3\epsilon t))) dt,\cdots
 \end{multline} 
 for continued exponential with Borel-Leroy transformation, \begin{multline}
           G_1 \equiv \log(g_1 \epsilon +1),\, G_2 \equiv \log(g_1 \epsilon \log(g_2 \epsilon+1)+1),\, G_3 \equiv \log(g_1 \epsilon \log(g_2 \epsilon \log(g_3 \epsilon +1)+1)+1),\\ G_4 \equiv \log(g_1 \epsilon \log(g_2 \epsilon \log(g_3 \epsilon\log(g_4 \epsilon +1) +1)+1)+1),\cdots\end{multline} for continued logarithm and \begin{equation}
      H_1\equiv\frac{h_0 }{h_1\epsilon+1},\,H_2\equiv\frac{h_0}{\frac{h_1\epsilon}{h_2\epsilon+1}+1},\,H_3\equiv\frac{h_0}{\frac{h_1\epsilon}{\frac{h_2\epsilon}{h_3\epsilon+1}+1}+1},\,H_4\equiv\frac{h_0}{\frac{h_1\epsilon}{\frac{h_2\epsilon}{\frac{h_3\epsilon}{h_4 \epsilon+1}+1}+1}+1},\cdots
  \end{equation} for continued fraction. \\ These transformed sequences are calculated for finding a numerical estimate from transformed variables $\{c_i\}$, $\{d_i\}$, $\{e_i\}$, $\{g_i\}$, $\{h_i\}$. These variables can be obtained as general expressions for any quantity $Q(\epsilon)$ by Taylor expansion of sequence at arbitrary order and from relations with coefficients $\{q_i\}$ of Eq.(1) such as (weighted averages of $\{q_i\}$)\begin{equation}
    \begin{gathered}
    q_0 = c_0 \hbox{e}^1,\,
    q_1 = -c_0c_1\hbox{e}^2,\,q_2 = c_0\hbox{e}^3{\left( c_1c_2 +\frac{3{c_1}^2}{2}\right)},\,
    q_3 = -c_0c_1 \hbox{e}^4 {\left(\frac{13{c_1 }^2}{6}+3c_1c_2+ \,c_2 \,c_3 +\frac{3{c_2 }^2 }{2} \, \right)},\, \cdots,
\end{gathered}
\end{equation} for continued exponential fraction \begin{equation}
    \begin{gathered}
    q_0 = d_0,\,
    q_1 = d_0 d_1,\,q_2 = d_0{\left(d_1 d_2 +\frac{{d_1 }^2 }{2}\right)},
    q_3 = d_0 d_1 {\left( d_2 \,d_3 +\frac{{d_2 }^2 }{2} +d_1d_2+\frac{{d_1 }^2 }{6} \right)},\, \cdots,
\end{gathered}
\end{equation} for continued exponential \begin{equation}
          q_1 = g_1,\,q_2 = g_1g_2,\,q_3 = g_1g_2g_3,\,q_4 = g_1g_2g_3g_4,\,q_5 = g_1g_2g_3g_4g_5,\,q_6 = g_1g_2g_3g_4g_5g_6,\cdots,
    \end{equation} for continued logarithm and \begin{equation}
     q_0=h_0,\,q_1=-h_0h_1,\,q_2=h_0h_1{\left(h_2+h_1 \right)},\,q_3=-h_0h_1 {\left(h_2 h_3 +{h_2}^2 +2h_1h_2 +{h_1}^2 \right)},\cdots
 \end{equation} for continued fraction. Solving relations in Eq.(14) for coefficients of Borel-Leroy transformed series $f(y)$ in Eq. (7) provides transformed variables $\{ e_i \}$ for continued exponential with Borel-Leroy transformation. While using the continued logarithmic function, $\left(g_i\epsilon\log(\cdots)+1\right)>0$ condition must be satisfied in every term of the sequence of Eq. (11), or the estimate becomes undefined. And, it is to be noted that this function is applicable only for quantities $Q(\epsilon)$ with $q_0=0$. \\ To attain the reliability of the estimates generated by these procedures, error calculation is essential. This is predicted by the principle of fastest apparent convergence, which measures differences of estimates at consecutive orders \cite{sixloop,ERROR}. The partial sums can be paired with the Shanks transformation for transformed sequences with convergence behaviours to produce accelerated convergence and assess their error \cite{bender1999advanced}. Shanks transformation for a convergent sequence $\{A_i\}$ is defined as \begin{equation}
  S(A_i) = \frac{A_{i+1}A_{i-1}-A_i^2}{A_{i+1}+A_{i-1}-2A_i},
 \end{equation} and iterated Shanks is $S^2(A_i) \equiv S(S(A_i))$. When $S^2(A_{i})$ is considered as prediction for $Q(\epsilon)$ the error is estimated from relation \cite{shalaby2020critical} \begin{equation}
     (|S(A_{i+1}) - S(A_{i})| + |S(A_{i+1}) - S^2(A_{i})|)/2.
 \end{equation} 
 Minimizing this error calculated from successive iterations is also helpful in determining the Borel-Leroy parameter $l$ or tuning parameter in Eq. (7).
    \subsubsection{ Critical exponents $\nu$ and $\omega$ of correlation length $\xi$}
\begin{figure}
    \centering
    \includegraphics[width=.6\linewidth, height=7cm]{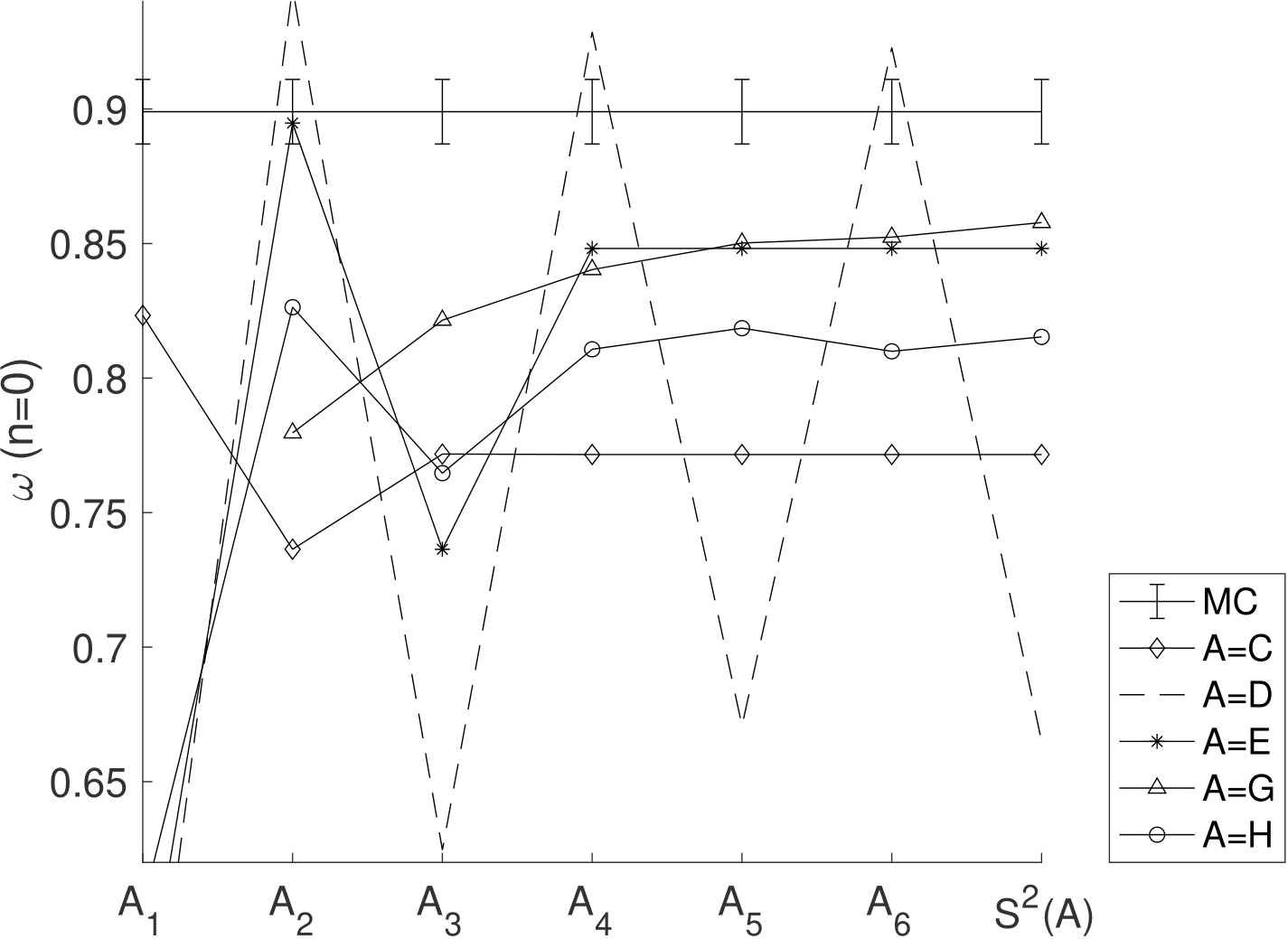}
    \caption{Estimates of $\omega$ at successive orders compared with MC result \cite{MCn0,shalaby2020critical} for self-avoiding walks model.}
    \label{n=0}
\end{figure}

   Theoretically, close to the critical point, the correlation length $\xi$ of the fluctuations associated with the field is the most important characteristic length scale. These fluctuations are responsible for the critical behaviour of all the thermodynamic quantities. The divergence of $\xi$ is controlled by the critical exponents $\nu$ and $\omega$ as
\begin{equation}
\xi(T) \sim |T-T_c|^{-\nu}(1+\hbox{const}.|T-T_c|^{\omega \nu}+\cdots). \end{equation}
For a $n$-component field, these critical exponents are derived as a power series of $\epsilon = (4-d)$. In our previous work, continued exponential and continued exponential fraction \cite{abhignan2020continued} were used to determine the exponent $\nu$, whereas the recent seven-loop perturbative expansion of $\omega$ for $n=0,1,2,3$ have the form of \cite{shalaby2020critical} 
\begin{subequations}
\begin{align}
    \omega &= \epsilon-0.65625\epsilon^2+1.8236\epsilon^3-6.2854\epsilon^4+26.873\epsilon^5-130.01\epsilon^6+692.10\epsilon^7, \\
    \omega &= \epsilon-0.62963\epsilon^2+1.6182\epsilon^3-5.2351\epsilon^4+20.750\epsilon^5-93.111\epsilon^6+458.74\epsilon^7, \\
    \omega &= \epsilon-0.60000\epsilon^2+1.4372\epsilon^3-4.4203\epsilon^4+16.374\epsilon^5-68.777\epsilon^6+316.48\epsilon^7, \\
    \omega &= \epsilon-0.57025\epsilon^2+1.2829\epsilon^3-3.7811\epsilon^4+13.182\epsilon^5-52.204\epsilon^6+226.02\epsilon^7,
    \end{align}
\end{subequations}
for $\epsilon\rightarrow0$ respectively.
Since this series does not have a zeroth order coefficient ($q_0=0$), continued exponential and continued exponential fraction could not be directly used to determine a reliable numerical value for exponent $\omega$, and so continued fraction was implemented \cite{abhignan2020continued}. Another way of handling this kind of perturbative expansion is perhaps by realizing the series as $\omega/\epsilon$ (this may not always work correctly for a divergent series \cite{bender1999advanced}). This is used for transformation through continued exponential fraction, continued exponential, continued exponential with Borel-Leroy transformation and continued fraction as defined in Eq.s (2), (3), (4) and (6), respectively. One can directly implement the transformation of $\omega$ using continued logarithm as defined in Eq. (5). \\In this manner, the numerical estimates of $\omega$ for $d=3$ ($\epsilon=1$), self-avoiding walks model ($n=0$) are obtained from sequences $\{C_i\}$, $\{D_i\}$, $\{E_i\}$, $\{G_i\}$, $\{H_i\}$ and their final estimate is interpolated from Shanks in Eq. (17) as \begin{subequations}
    \begin{align}
C_1&=0.82327,C_2=0.73638,C_3=0.77176,C_4=0.77159,C_5=0.77158,C_6=0.77158,S^2(C_4)=0.77158, \\
D_1&=0.51879,D_2=0.94498,D_3=0.62464,D_4=0.92862,D_5=0.67078,D_6=0.92284,S^2(D_4)=0.665(71), \\
E_1&=0.55938,E_2=0.89477,E_3=0.73637,E_4=0.84816,E_5=0.84818,E_6=0.84816,S^2(E_4)=0.84817, \\
G_2&=-2.6906,G_3=0.77963,G_4=0.82150,G_5=0.84032,G_6=0.85018,G_7=0.85237,S^2(G_5)=0.8578(64),
\\
H_1&=0.60377,H_2=0.82633,H_3=0.76467,H_4=0.81070,H_5=0.81854,H_6=0.80994,S^2(H_4)=0.8153(33). 
 \end{align}
\end{subequations}
These estimates are illustrated in Fig. \ref{n=0} and compared with the most reliable MC result $\omega=0.899(12)$ \cite{MCn0,shalaby2020critical}. The error for final estimates is evaluated from Eq. (18). For continued exponential with Borel-Leroy transform, by tuning the parameter $l$, the plot for a final estimate $S^2(E_4)$ for $l \in [0.5,2]$ with error bars showing $(|S(E_{5}) - S(E_{4})| + |S(E_{5}) - S^2(E_{4})|)/2$ is illustrated in Fig. 2(a). As it is observed, the final estimate is sensitive to the tuning parameter and taken at $l=1.43$, where the prediction is most precise with $\omega=0.84817$. All estimates undershoot, while the continued logarithm estimate $\omega=0.8578(64)$ is most comparable to the MC value. However, when compared with recent resummation studies, the continued function with Borel-Leroy transform estimate, and continued logarithm estimate are most compatible with previous predictions from hypergeometric-Meijer resummation (HM) \cite{shalaby2020critical} (seven-loop) where $\omega=0.8484(17)$, Borel with conformal mapping calculations (BCM) \cite{sixloop} (six-loop) where $\omega=0.841(13)$ and self-consistent resummation algorithm (SC) \cite{selfconsistent} where $\omega=0.846(15)$. It is also interesting to observe that oscillating sequence from continued exponential envelops the region of convergence from different approaches. 
\begin{figure}[hb]
\centering

\begin{subfigure}{0.5\textwidth}
\includegraphics[width=1.2\linewidth, height=8cm]{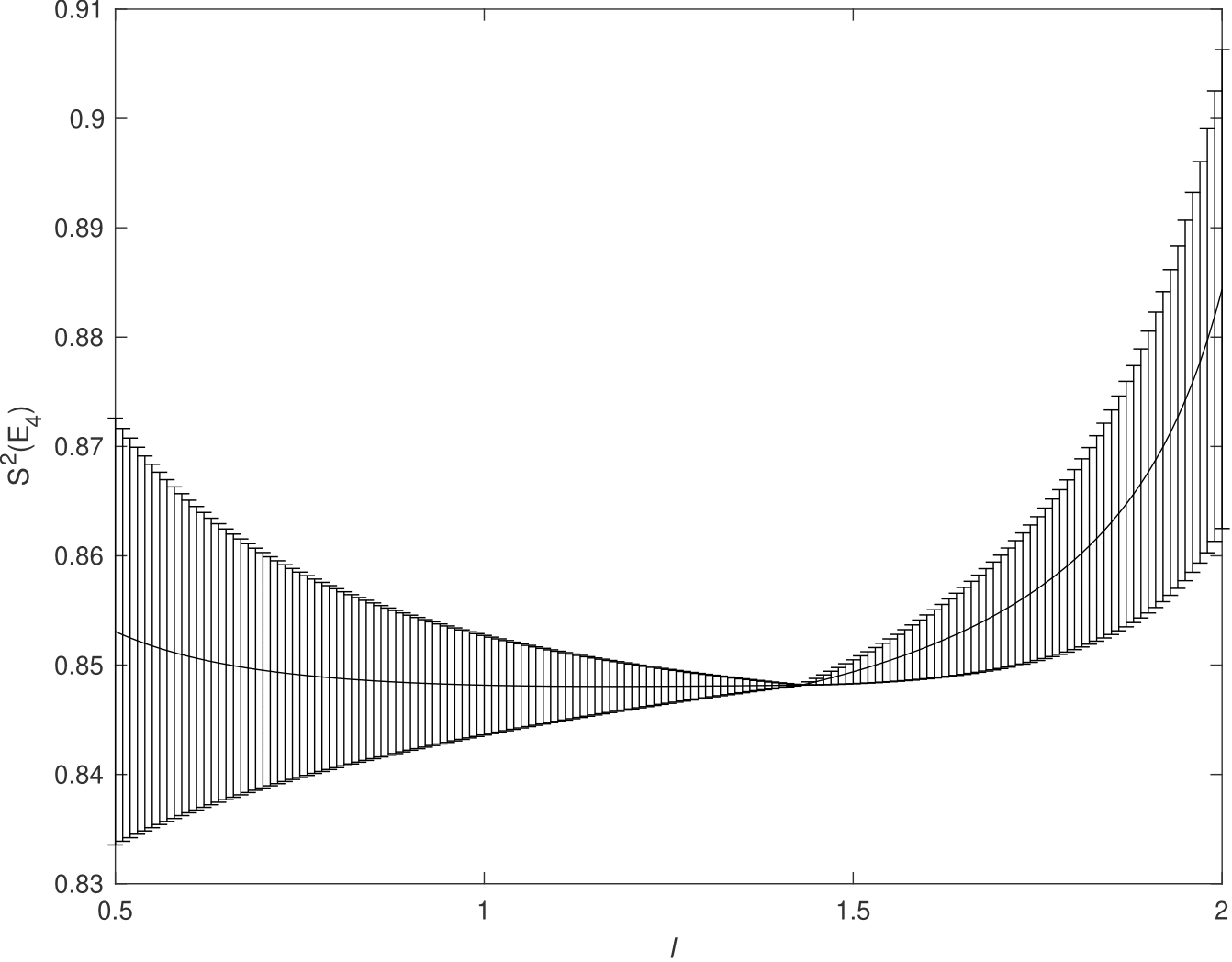}
\caption{$S^2(E_4)$ vs $l,\,(n=0)$ for $l \in [0.5,2]$.}

\end{subfigure}
\begin{subfigure}{0.5\textwidth}
\includegraphics[width=1.2\linewidth, height=8cm]{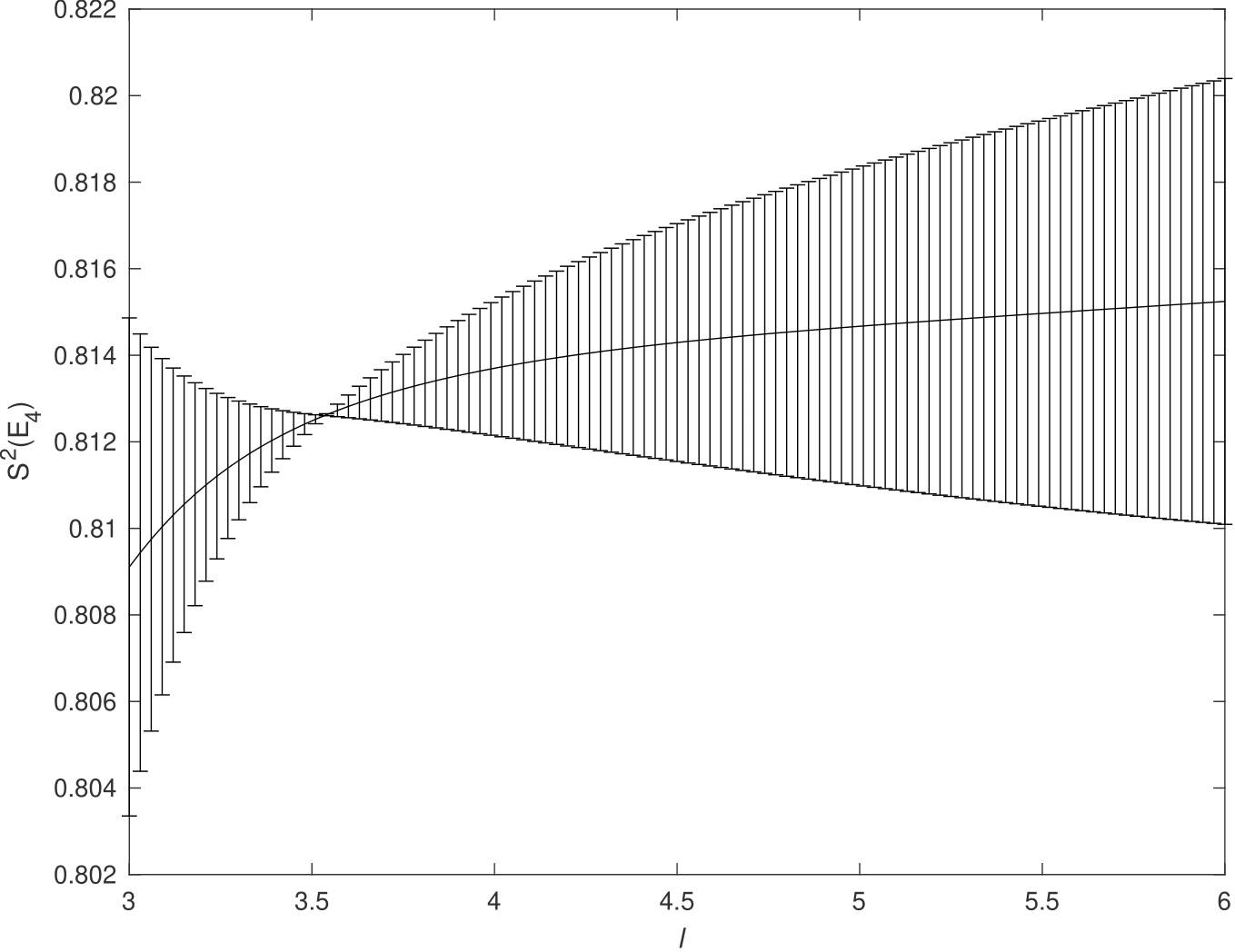} 
\caption{$S^2(E_4)$ vs $l,\,(n=1)$ for $l \in [3,6]$.}

\end{subfigure}

\caption{The estimate of $\omega$ derived from $S^2(E_4)$ vs shift parameter $l$ is plotted, with the error bars showing the value of $(|S(E_{5}) - S(E_{4})| + |S(E_{5}) - S^2(E_{4})|)/2$.}

\end{figure} 
\\The numerical estimates of $\omega$ for $d=3$, Ising-like model ($n=1$) are obtained from sequences
\begin{subequations}
    \begin{align}
C_1&=0.82856,C_2=0.73919,C_3=0.77329,C_4=0.77278,C_5=0.77269,C_6=0.77272,S^2(C_4)=0.77270(2), \\
D_1&=0.53279,D_2=0.93612,D_3=0.63803,D_4=0.91397,D_5=0.68127,D_6=0.90431,S^2(D_4)=0.741(30),  \\
E_1&=0.55458,E_2=0.90540,E_3=0.70533,E_4=0.85741,E_5=0.79393,E_6=0.82036,S^2(E_4)=0.81259(2), \\
G_2&=-4.9986,G_3=0.75617,G_4=0.77396,G_5=0.79920,G_6=0.80725,G_7=0.81013,S^2(G_5)=0.81174(36),
\\
H_1&=0.61364,H_2=0.82364,H_3=0.76342,H_4=0.80579,H_5=0.80533,H_6=0.80578,S^2(H_4)=0.80556(11).
 \end{align}
\end{subequations} These estimates are illustrated in Fig. \ref{n=1} and compared with MC result $\omega=0.832(6)$ \cite{MCHAS}. To illustrate the repeatability of behaviour in continued exponential with Borel-Leroy transform, we calculate $S^2(E_4)$ here for varying Borel-Leroy parameter $l$ and plot it in Fig. 2(b). The obtained precise prediction is $\omega=0.81259(2)$ at $l=3.53$. This value and continued logarithm estimate $\omega=0.81174(36)$ are most comparable with the MC result, while other estimates undershoot. The continued logarithm estimate and continued exponential with Borel-Leroy transform estimate are also compatible with recent HM prediction $\omega=0.82311(50)$, BCM calculation $\omega=0.820(7)$ and SC algorithm $\omega=0.827(13)$. \begin{figure}
    \centering
    \includegraphics[width=.6\linewidth, height=7cm]{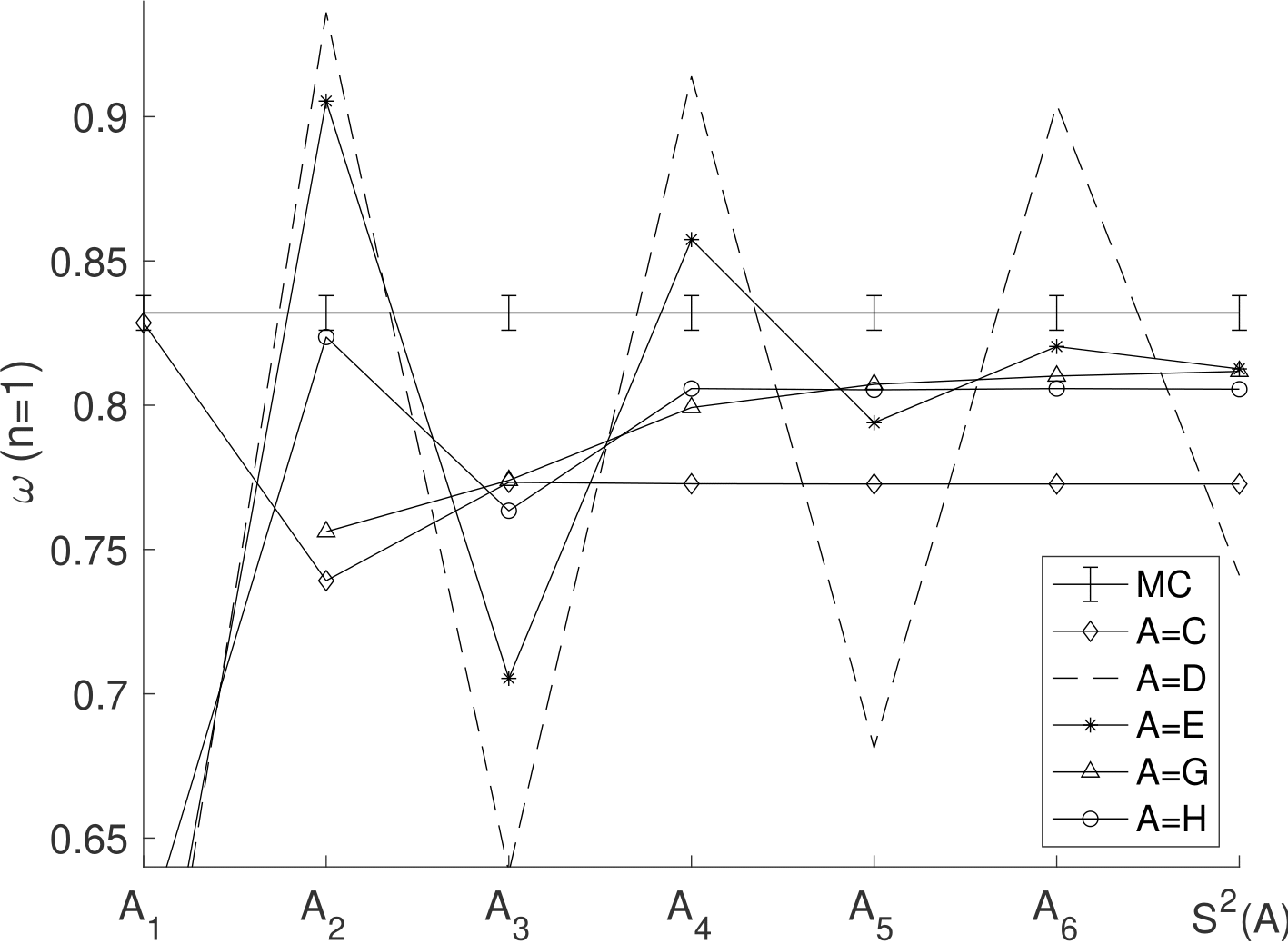}
    \caption{Estimates of $\omega$ at successive orders compared with MC result \cite{MCHAS} for Ising-like model.}
    \label{n=1}
\end{figure} \\The numerical estimates of $\omega$ for $d=3$, $XY$ universality class ($n=2$) are obtained from sequences
\begin{subequations}
    \begin{align}
C_1&=0.83459,C_2=0.74368,C_3=0.77630,C_4=0.77522,C_5=0.77475,C_6=0.77498,S^2(C_4)=0.77472(34), \\
D_1&=0.54881,D_2=0.92884,D_3=0.65044,D_4=0.89975,D_5=0.68808,D_6=0.88433,S^2(D_4)=0.774(10),  \\
E_1&=0.58735,E_2=0.87944,E_3=0.74371,E_4=0.81789,E_5=0.81785,E_6=0.74371,S^2(E_4)=0.81789(2), \\
G_2&=-2.4804,G_3=0.73100,G_4=0.72010,G_5=0.75477,G_6=0.75999,G_7=0.76383,S^2(G_5)=0.81174(36), 
\\
H_1&=0.62500,H_2=0.82329,H_3=0.76388,H_4=0.80231,H_5=0.79408,H_6=0.80029,S^2(H_4)=0.7983(14).
 \end{align}
\end{subequations} These estimates are illustrated in Fig. \ref{n=2} and compared with MC result  $\omega=0.789(4)$ \cite{mcn2}. The continued fraction estimate $\omega=0.7983(14)$ is most comparable with the MC value, while other estimates undershoot or overshoot. The continued function with Borel-Leroy transform estimate $\omega=0.81789$ ($l=1.26$) and continued logarithm estimate $\omega=0.81174(36)$ is most compatible with CB result $\omega=0.811(10)$ \cite{Echeverri2016}, HM prediction $\omega=0.789(13)$ and BCM calculation $\omega=0.804(3)$.  \begin{figure}
    \centering
    \includegraphics[width=.6\linewidth, height=7cm]{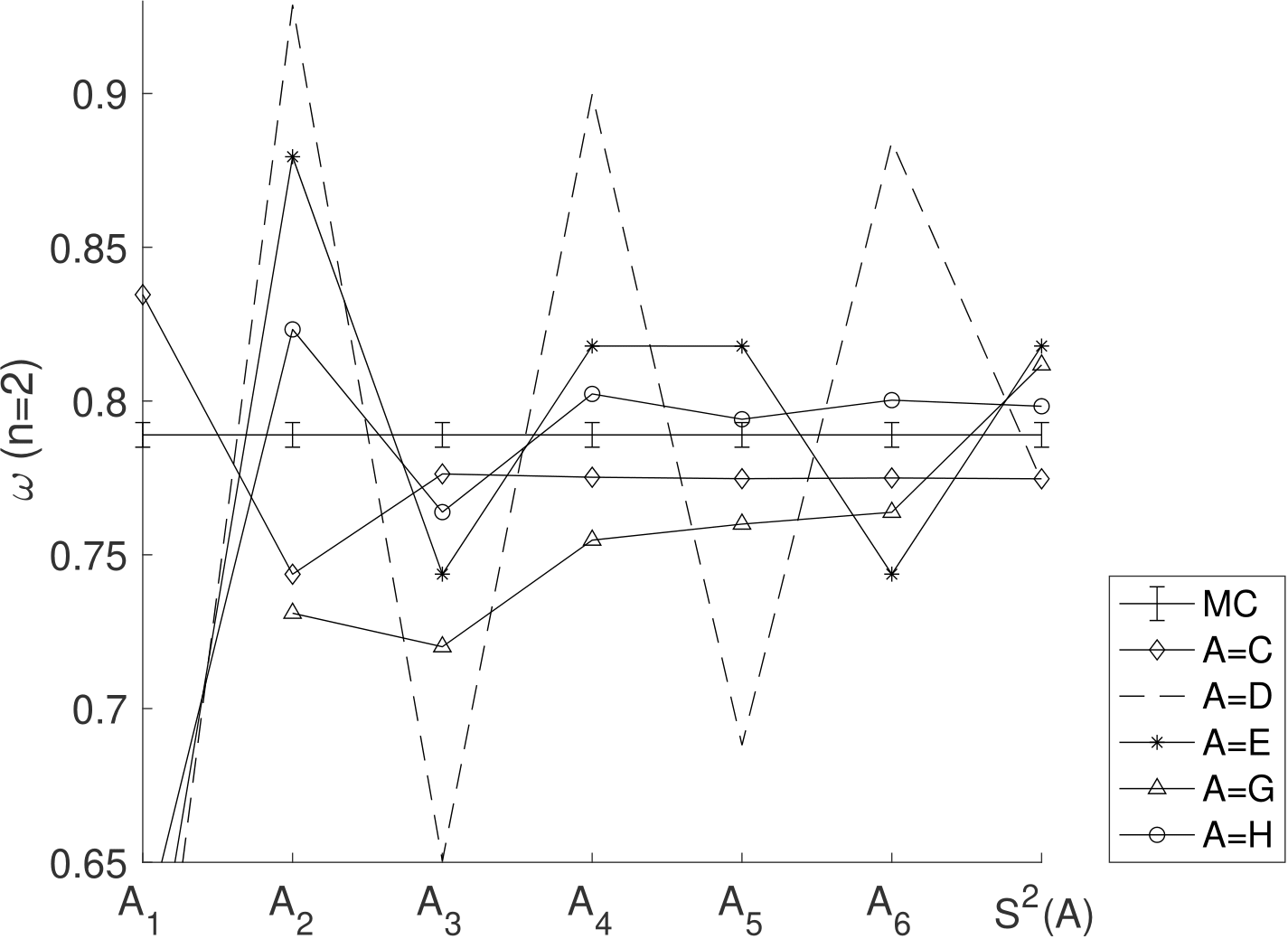}
    \caption{Estimates of $\omega$ at successive orders compared with MC result \cite{mcn2} for $XY$ universality class.}
    \label{n=2}
\end{figure} \\The numerical estimate of $\omega$ for $d=3$, Heisenberg universality class ($n=3$) are obtained from sequences
\begin{subequations}
    \begin{align}
C_1&=0.84080,C_2=0.74909,C_3=0.78036,C_4=0.77855,C_5=0.77572,C_6=0.77784,S^2(C_4)=0.7807(52), \\
D_1&=0.56538,D_2=0.92316,D_3=0.66270,D_4=0.88691,D_5=0.69486,D_6=0.86466,S^2(D_4)=0.7831(17),  \\
E_1&=0.60287,E_2=0.87481,E_3=0.74859,E_4=0.80427,E_5=0.804425,E_6=0.80427,S^2(E_4)=0.80435(4), \\
G_2&=-1.8614,G_3=0.70588,G_4=0.66200,G_5=0.70910,G_6=0.71041,G_7=0.71550,S^2(G_5)=0.70877(96), 
\\
H_1&=0.63684,H_2=0.82452,H_3=0.76614,H_4=0.80025,H_5=0.78725,H_6=0.79218,S^2(H_4)=0.79083.
 \end{align}
\end{subequations} These estimates are illustrated in Fig. \ref{n=3} and compared with MC result $\omega=0.773$ \cite{mcn3}. The continued exponential estimate $\omega=0.7831(17)$, continued exponential fraction estimate $\omega=0.7807(52)$ are comparable with the MC value and similarly continued fraction estimate $\omega=0.79083$, continued exponential with Borel-Leroy transform estimate $\omega=0.80435(4)$ ($l=1.16$) are most compatible with predictions from BCM calculation $\omega=0.795(7)$ and SC resummation algorithm $\omega=0.794(4)$.
\begin{figure}
    \centering
    \includegraphics[width=.6\linewidth, height=7cm]{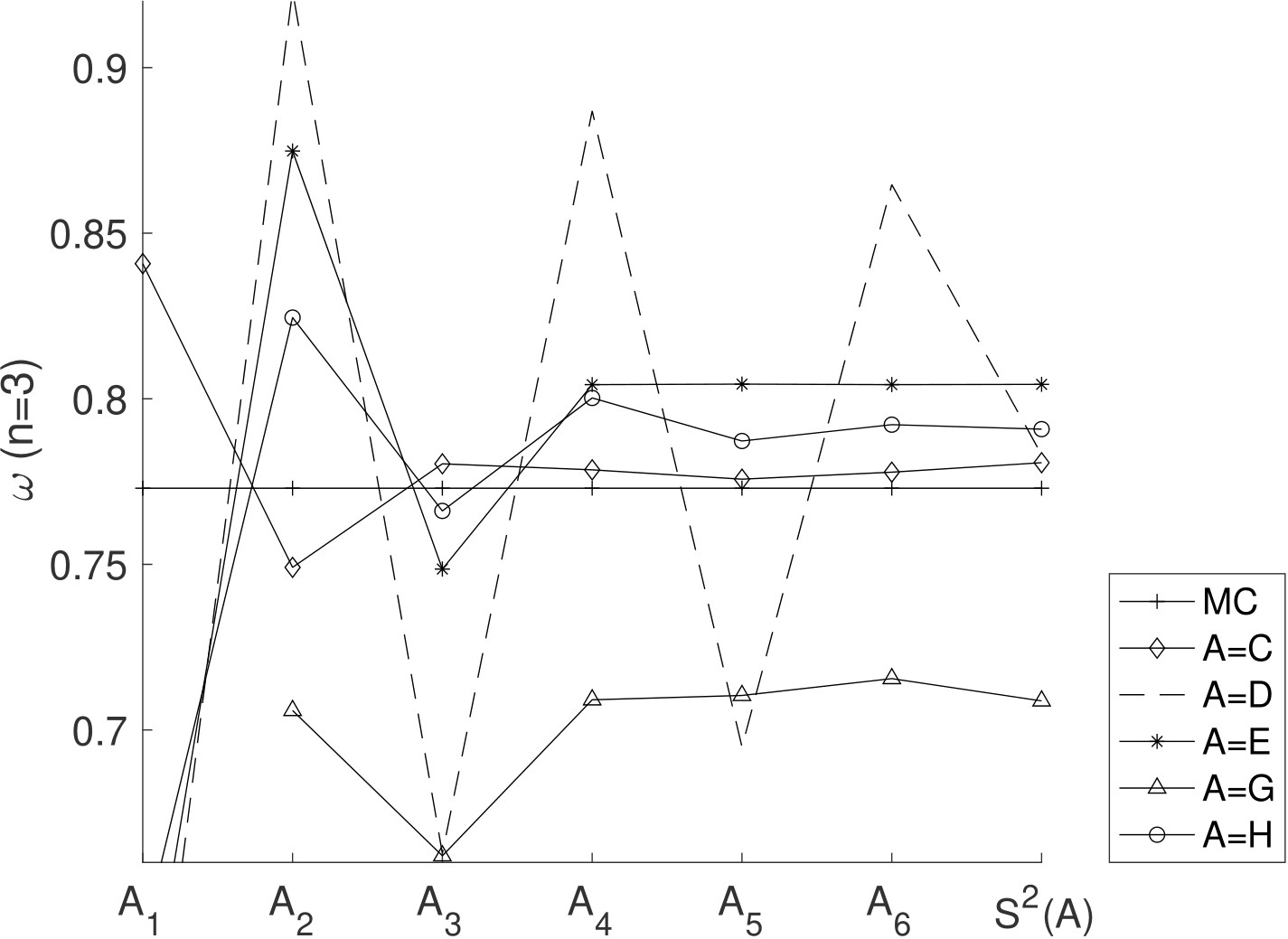}
    \caption{Estimates of $\omega$ at successive orders compared with MC result \cite{mcn3} for Heisenberg universality class.}
    \label{n=3}
\end{figure} 

Similarly, we study $\omega$ for $d=2$ ($\epsilon=2$) systems. These results are interesting since previous resummation studies of RG functions could not predict reliable estimates for two-dimensional systems \cite{Eckmann1975,Orlov2000,Calabrese_2000} due to non-analyticity of $\beta$-functions around the fixed point. We obtain numerical estimates of $\omega$ for $d=2$ self-avoiding walks model (Eq. (20a)) from the sequences at consecutive orders as \begin{subequations}
    \begin{align}
C_1&=1.4442,C_2=1.3059,C_3=1.3685,C_4=1.3682,C_5=1.3682,C_6=1.3682,S^2(C_4)=1.3682, \\
D_1&=0.53829,D_2=1.9806,D_3=0.60286,D_4=1.9793,D_5=0.61259,D_6=1.9792,S^2(D_4)=1.276(11),  \\
E_1&=0.68393,E_2=1.8792,E_3=1.0471,E_4=1.7958,E_5=1.7542,E_6=1.0583,S^2(E_4)=1.805(24), \\
G_2&=1.8597,G_3=1.6694,G_4=1.8057,G_5=1.8065,G_6=1.8152,G_7=1.8154,S^2(G_5)=1.8064(93),
\\
H_1&=0.86486,H_2=1.5997,H_3=1.3194,H_4=1.5533,H_5=1.6082,H_6=1.5509,S^2(H_4)=1.589(27).
 \end{align}
\end{subequations} These estimates are illustrated in Fig. \ref{n=0,d=2} and compared with exact result from lattice models, $\omega=2$ \cite{twodimension,Caracciolo2005}. Continued with Borel-Leroy transform estimate $\omega=1.805(24)$ ($l=1.75$), continued logarithm estimate $\omega=1.8064(93)$ are comparable with the exact result, HM resummation $\omega=1.96(46)$ and BCM prediction $\omega=1.90(25)$. \begin{figure}
    \centering
    \includegraphics[width=.6\linewidth, height=7cm]{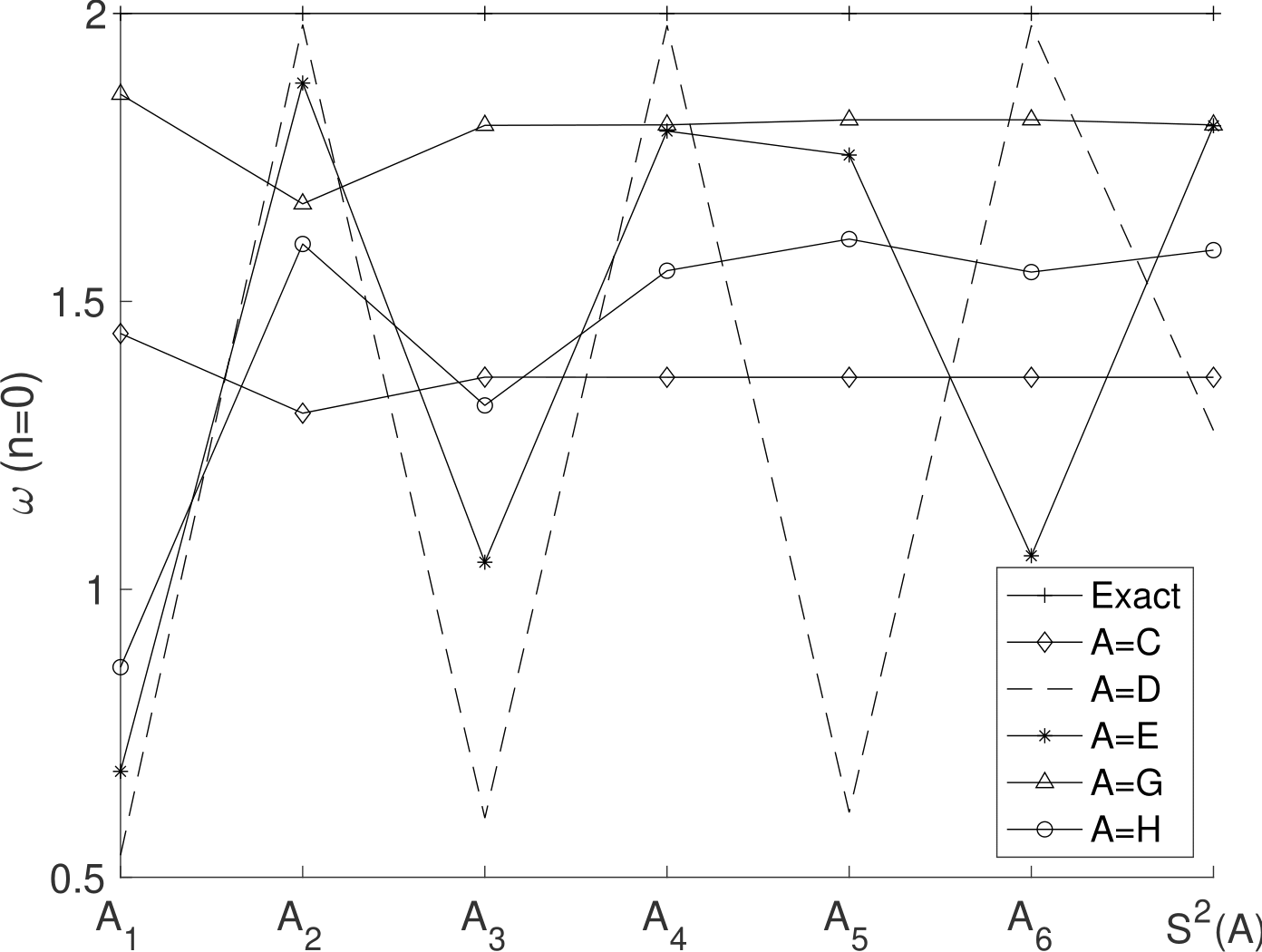}
    \caption{Estimates of $\omega$ at successive orders for self-avoiding walks model compared with exact result \cite{twodimension,Caracciolo2005} in two-dimensional system.}
    \label{n=0,d=2}
\end{figure}
\\We obtain numerical estimates of $\omega$ for $d=2$ Ising-like model from the sequences as \begin{subequations}
    \begin{align}
C_1&=1.4573,C_2=1.3113,C_3=1.3725,C_4=1.3717,C_5=1.3716,C_6=1.3716,S^2(C_4)=1.3716(1), \\
D_1&=0.56773,D_2=1.9725,D_3=0.64065,D_4=1.9696,D_5=0.65243,D_6=1.9693,S^2(D_4)=1.2981(78),  \\
E_1&=0.72905,E_2=1.8469,E_3=1.1097,E_4=1.7273,E_5=1.7283,E_6=1.7273,S^2(E_4)=1.7278(3), \\
G_2&=1.8732,G_3=1.6453,G_4=1.7843,G_5=1.7842,G_6=1.7936,G_7=1.7938,S^2(G_5)=1.7842(95),
\\
H_1&=0.88525,H_2=1.5898,H_3=1.3127,H_4=1.5348,H_5=1.5316,H_6=1.5348,S^2(H_4)=1.5333(8).
 \end{align}
\end{subequations} 
These estimates are illustrated in Fig. \ref{n=1,d=2} and compared with the exact result from the lattice model, $\omega=1.75$ \cite{Calabrese_2000}. For this model, the estimate from continued exponential with Borel-Leroy transform $\omega=1.7278(3)$ ($l=1.36$) and continued logarithm estimate $\omega=1.7842(95)$ are compatible with the exact value. Similarly, the HM prediction $\omega=1.71(10)$ and BCM calculation $\omega=1.71(9)$ for the Ising-like model illustrate that resummation of $\epsilon$-expansions gives better estimates than the resummation of the coupling-series \cite{kleinert}. It is noted here that while procedures such as BCM and HM implement, large-order asymptotic information of RG functions, our methods implement only the lower-order information as described. Other exponents related to scaling exponent $\nu$ can be approximately estimated from the Gaussian model (mean-field theory). In contrast, the measurement of subleading exponent $\omega$ completely requires the corrections from perturbative RG and is important to understand the relevant directions in RG flows. Hence the procedures where only the corrections are used without external parameters to measure the exponent $\omega$ are more reliable.
\begin{figure}
    \centering
    \includegraphics[width=.6\linewidth, height=7cm]{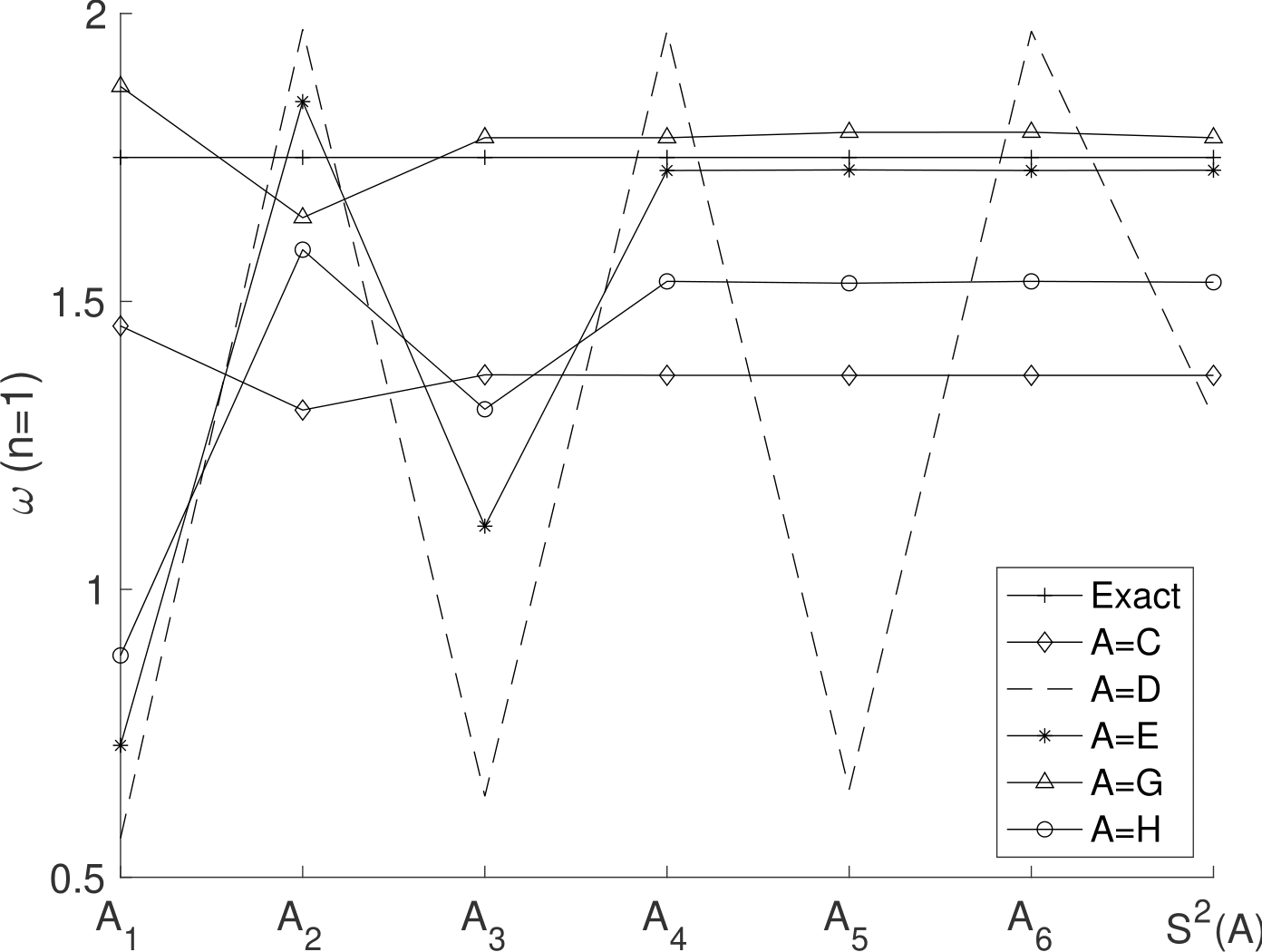}
    \caption{Estimates of $\omega$ at successive orders for Ising-like model compared with exact result \cite{Calabrese_2000} in two dimensional system.}
    \label{n=1,d=2}
\end{figure}
     \\A more stringent check for these procedures would be to obtain the most accurately measured result from the microgravity experiment for superfluid helium where specific heat exponent $\alpha_{XY}=-0.0127(3)$ \cite{LIpa2003}. Using the slowly converging seven-loop $\epsilon$ expansion \cite{shalaby2020critical} of \begin{equation}
        \nu_{XY}=2.0000-0.40000\epsilon - 0.14000\epsilon^2 + 0.12244\epsilon^3 -0.30473\epsilon^4 + 0.87924\epsilon^5 - 3.1030\epsilon^6 + 12.419\epsilon^7,
    \end{equation} we obtain the sequences directly for continued exponential fraction, continued exponential, continued exponential with Borel-Leroy transform, continued fraction and estimates for $\nu_{XY}$ in $d=3$ at consecutive orders as
\begin{subequations}
    \begin{align}
C_1&=0.53547,C_2=0.61992,C_3=0.71029,C_4=0.66851,C_5=0.67366,C_6=0.67157,C_7=0.67161,\nonumber \\ & \hspace{128mm} S^2(C_5)=0.67070(74), \\
D_1&=0.61070,D_2=0.68421,D_3=0.64135,D_4=0.67758,D_5=0.65436,D_6=0.67576,D_7=0.65963,\nonumber \\ & \hspace{128mm} S^2(D_5)=0.6677(11), \\
E_1&=0.61018,E_2=0.68298,E_3=0.64144,E_4=0.67610,E_5=0.65569,E_6=0.67395,E_7=0.66219,\nonumber \\ & \hspace{128mm} S^2(D_5)=0.6704(25), \\
H_1&=0.60000,H_2=0.72222,H_3=0.64474,H_4=0.67334,H_5=0.66396,H_6=0.67012,H_7=66944,\nonumber \\ & \hspace{128mm} S^2(H_5)=0.6617(48). 
 \end{align}
\end{subequations} 
We compare these estimates with microgravity experimental value (exp) $\nu_{XY}=0.6709(1)$ \cite{LIpa2003} and most reliable MC result $\nu_{XY}=0.67169(7)$ \cite{mcn2} in Fig. \ref{nuxy}. \begin{figure}[ht]
\centering
\begin{subfigure}{0.5\textwidth}
\includegraphics[width=1.2\linewidth, height=8cm]{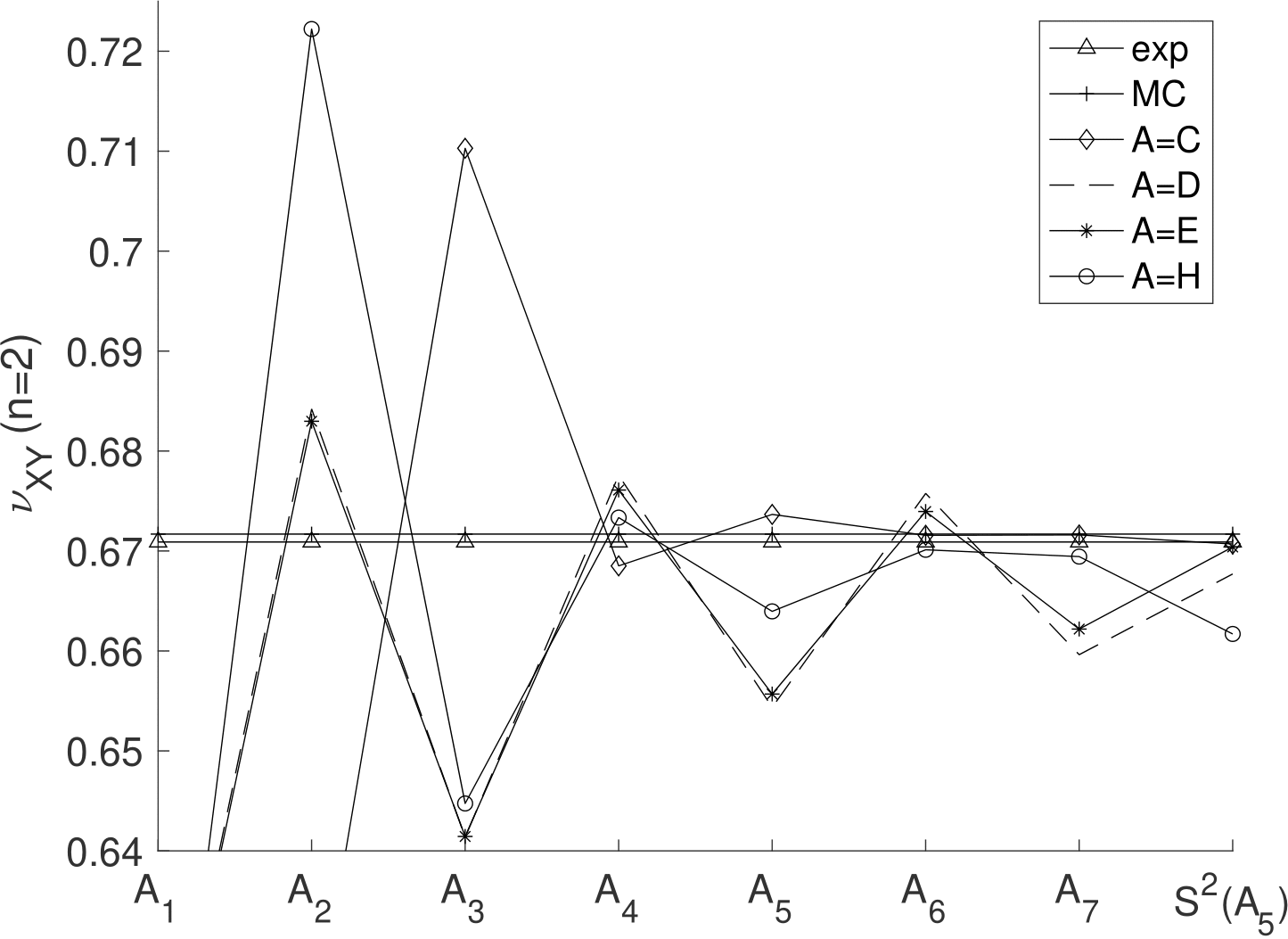} 
\caption{Oscillating convergence of $\nu_{XY}$ from continued functions.}

\end{subfigure}
\begin{subfigure}{0.5\textwidth}
\includegraphics[width=1.2\linewidth, height=8cm]{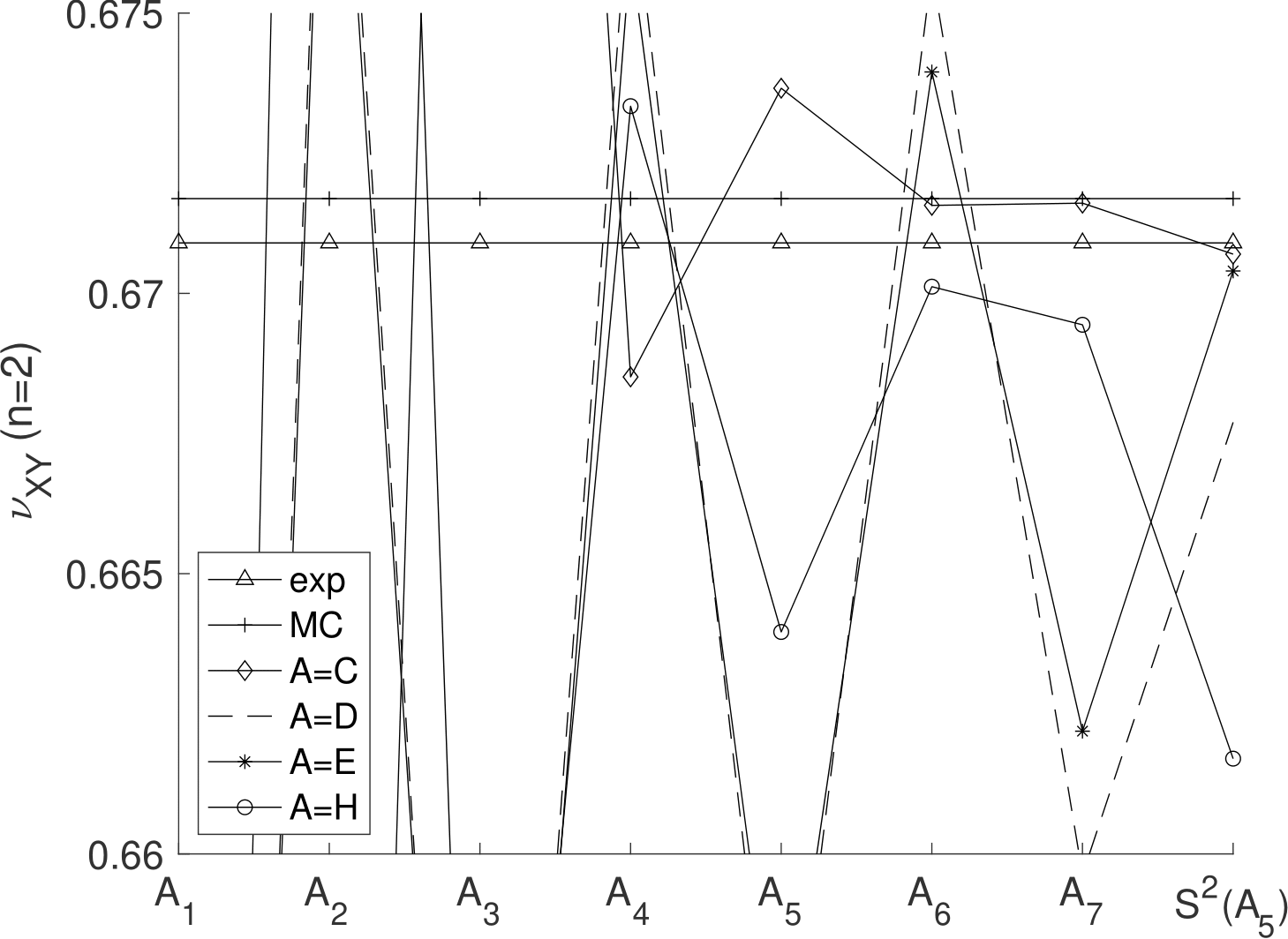}
\caption{Seven-loop resummation of $\nu_{XY}$.}

\end{subfigure}

\caption{The estimates of $\nu_{XY}$ compared with precise experimental value \cite{LIpa2003} and MC result \cite{mcn2}.}
\label{nuxy}
\end{figure} The oscillating sequences converging towards these precise values can be visualized in Fig. \ref{nuxy}(a). Using the continued exponential with Borel-Leroy transform, we obtain the estimate $\nu_{XY}=0.6704(25)$ at $l=22.51$ (Fig. 9), where it is observed that $\nu_{XY} \in [0.669,0.677]$ for $l \in [10,25]$. 

\begin{figure}
    \centering
    \includegraphics[width=.6\linewidth, height=7cm]{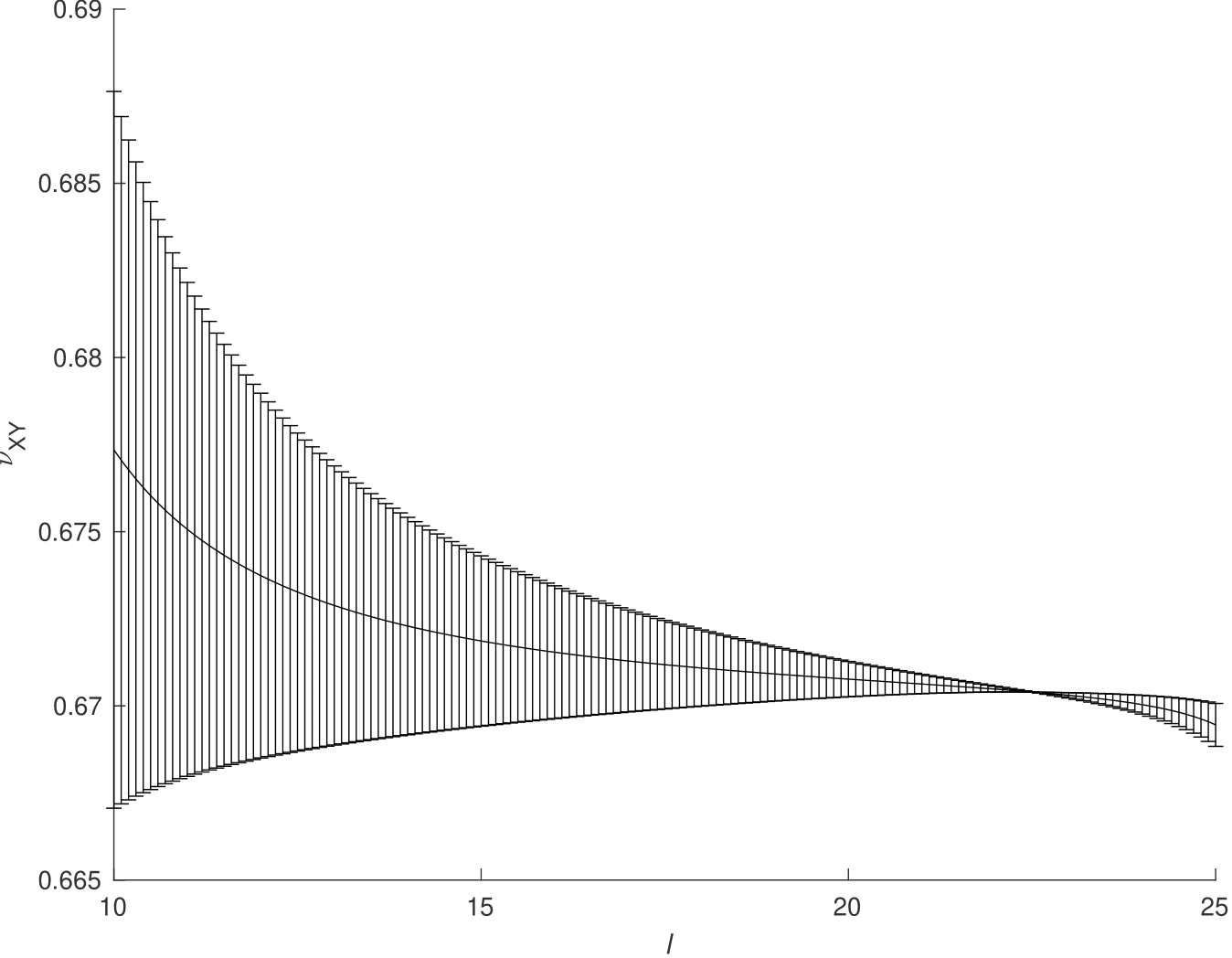}
    \caption{The estimate of $\nu_{XY}$ vs shift parameter $l$ is plotted, with the error bars showing the relation for error.}
\end{figure}
Further using Josephson's identity, $\alpha=2-d \nu$, we obtain estimates for continued exponential fraction, continued exponential, continued exponential with Borel-Leroy transform and continued fraction as $\alpha_{XY}=-0.0121(22)$, $\alpha_{XY}=-0.017(20)$, $\alpha_{XY}=-0.0112(76)$ and $\alpha_{XY}=0.015(14)$, respectively. The initial three estimates and recent resummation of seven-loop RG $\alpha_{XY}=-0.0123(11)$ (HM) seem more compatible with the precise experimental value. However, the significant errors in these predictions from RG are concerning since they cannot completely address the mismatch of predictions from MC and experimental value \cite{lambda} which can be distinctly seen in Fig. \ref{nuxy}(b). Eight-loop RG functions may help in resolving these issues. 
\subsection{Lattice Ising model ($n=1$)}
Considering the microscopic degrees of freedom, the discrete lattice Ising model provides the same statistical description to describe the nature of continuous phase transitions in $O(1)$ $\phi^4$ field models. The partition function of the simplest one-dimensional Ising model \cite{kardar2007statistical} is, 

\begin{equation}
    Z = \sum_{\{\sigma_i\}} \exp{  \left[ \sum_{<i,j>} B(\sigma_i,\sigma_j)\right]} = \sum_{\{\sigma_i\}} \exp{\left[ K \sum_{<i,j>} \sigma_i \sigma_j + \frac{h}{2} \sum_{<i,j>} (\sigma_i + \sigma_j)  \right]}.
\end{equation}

where ${\sigma_i}=\pm1$ is the spin at each lattice site $i$. $B(\sigma_i,\sigma_j)$ is the energy per bond between two nearest neighbour lattice sites $i$ and $j$. Here  $K=J/k_B T$, $J$ is the nearest neighbour coupling constant, and $h$ is the external magnetic field. The partition function just over the nearest neighbours is taken as
\begin{equation}
      Z = \sum_{\{\sigma_i\}}^{N'} \exp{ \left[ \sum_{<i,i+1>}^{N'} B(\sigma_i,\sigma_{i+1})\right]}. \label{Ising}
\end{equation} $\sum_{\{\sigma_i\}}^{N'}$ indicates summing over all possible $2^{N'}$ configurations of $N'$ spins and $\sum_{<i,i+1>}$ indicates summation over all nearest neighbour pairs. Though no phase transition exists on this one-dimensional model, extension to two dimensions led to interesting analytical conclusions based on Kramers-Wannier duality \cite{PhysRevST} in the seminal work by Onsager \cite{onsager}. This is used to study ordering in paramagnetic-ferromagnetic transitions.
 \subsubsection{Low temperature expansions}
 \begin{figure}[!htp]
\centering
\begin{subfigure}{0.5\textwidth}
\includegraphics[width=1.2\linewidth, height=8cm]{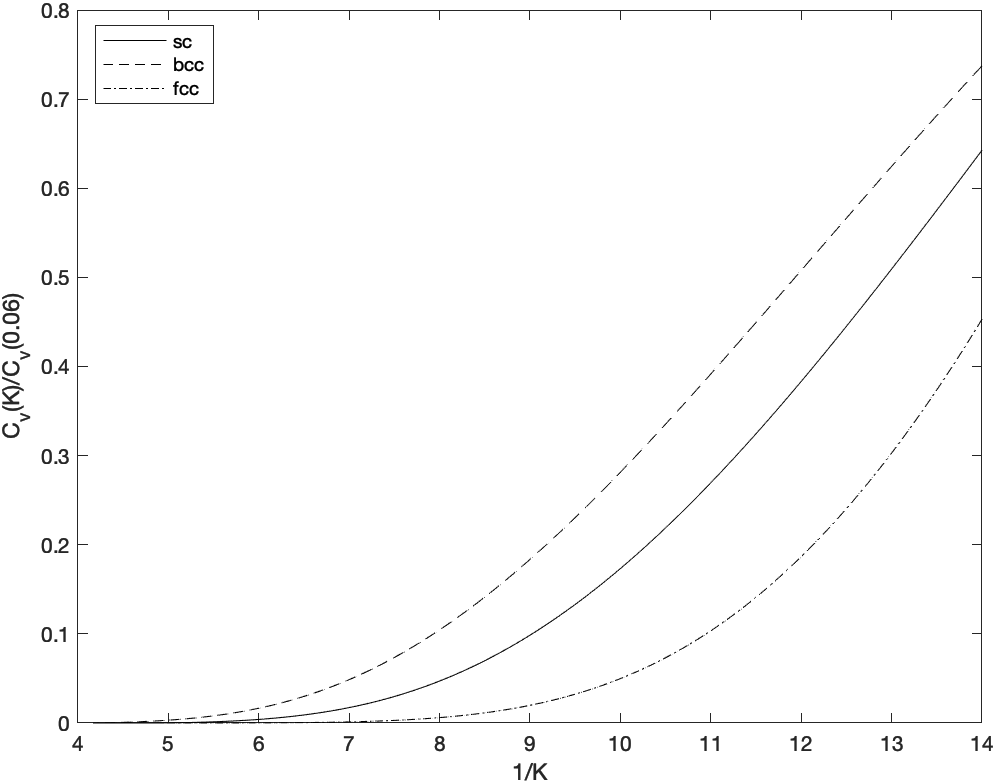} 
\caption{Comparing $C_v$ vs $1/K$ values for sc, bcc and fcc.}

\end{subfigure}
\begin{subfigure}{0.5\textwidth}
\includegraphics[width=1.2\linewidth, height=8cm]{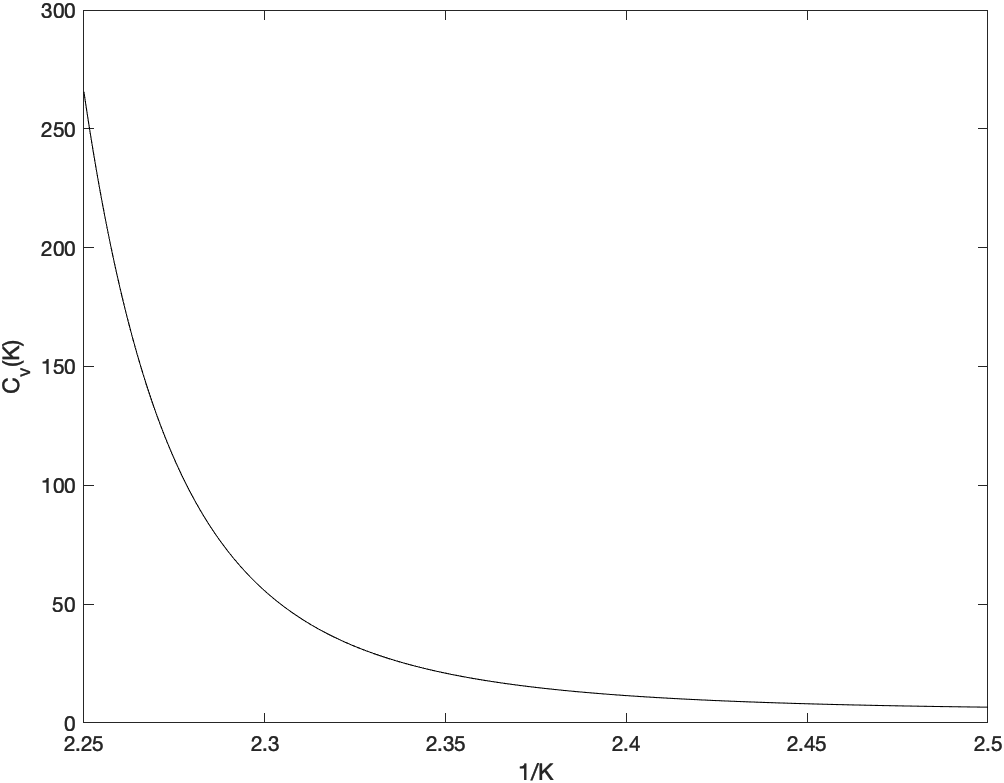}
\caption{$C_v$ vs $1/K$ values for sq.}

\end{subfigure}

\caption{Illustrating the behaviour of $C_v$ vs $1/K$ for $d=3$ and $d=2$ lattice around $1/K_c$.}

\end{figure}
Different perspectives are used for solving this partition function $Z$ of two and three-dimensional Ising models \cite{domb1960}. A diagrammatic approach was used to capture the co-existence of different phases in the vicinity of critical points using the method of low-temperature expansions. The partition function $Z$ was derived by studying excitation and interactions among the excitation around the most stable configuration at $T\rightarrow0$. These series expansions are divergent, and so initially, Pad\'e approximants were applied by Baker to obtain an analytic continuation \cite{Baker1963,domb1970}. Similarly, we use continued exponential to study the extensive quantity, specific heat $C_v$ derived from such low-temperature expansions in the factors of $u=\exp{(-4K)}$. The critical exponent $\alpha$ can be derived by studying the behaviour of  $C_v$ at constant volume near the critical temperature as $C_v \sim |T-T_c|^{-\alpha}$. \\Here we study the behaviour of $C_v(K)$ \cite{Baker1963} close to the critical point $K_c$ ($\sim 1/T_c$) for the $d=2$ simple quadratic lattice (sq) where \begin{equation}
   C_v(K)/K^2 = 64u^2+288u^3+1152u^4+4800u^5+21504u^6+101920u^7+502016u^8+2538432u^9+13078720u^{10}+68344496u^{11},
\end{equation}  $d=3$ simple cubic lattice (sc) where \begin{equation}
    C_v(K)/K^2 = 144u^3+1200u^5-2016u^6+11760u^7-33792u^8+135216u^9-448800u^{10}+1643664u^{11}-5671872^{12}, 
\end{equation} $d=3$ body-centred cubic lattice (bcc) where
\begin{equation}
    C_v(K)/K^2 = 256u^{4}+3136u^{7} -4608u^{8} +4480u^{10} - 123904u^{11}+111360u^{12} +551616u^{13} - 2464896u^{14}+4190400u^{15},
\end{equation}
and $d=3$ face-centered cubic lattice (fcc) where \begin{equation}
   C_v(K)/K^2 = 576u^6+11616u^{11}-14976u^{12}+28800u^{15}+172032u^{16}-554880u^{17}+374976u^{18}+138624^{19}+787200u^{20}.
   \end{equation} \\The Taylor expressions around $K=0$ for these expressions of $C_v(K)/K^2$ are recast into continued exponential (Eq. (3)) up to the ninth order such as \begin{multline} 
    \hbox{sq:}\,84593392\exp(-43.044K\exp(-0.0557K\exp(2.0512K\exp(0.9789K\exp(1.0873K\exp(1.1605K\\\exp(0.8735K\exp(0.2115K\exp(-9.9314K)))))))))
\end{multline}
\begin{multline}
    \hbox{sc:}\,-54793248\exp(-56.9186K\exp(0.0277K\exp(1.3411K\exp(2.2139K\exp(4.3708K\exp(1.5199K\\\exp(0.069K\exp(-57.104K\exp(34.032K)))))))))
\end{multline}
\begin{multline}
    \hbox{bcc:}\,-34418496\exp(-68.824K\exp(0.03089K\exp(2.0647K\exp(2.7629K\exp(5.6212K\exp(1.9136K\\\exp(-0.6834K\exp(14.292K\exp(-1.2633K)))))))))
\end{multline}
\begin{multline}
    \hbox{fcc:}\,4366089\exp(-96.531K\exp(0.2539K\exp(1.1231K\exp(-11.0745K\exp(25.0895K\exp(6.7036K\\\exp(13.5554K\exp(6.4608K\exp(15.9134K)))))))))
\end{multline} assuming that low-temperature expansions around $K=0$ are sufficient to capture the nature of singular behaviour. To illustrate the behaviour of expressions of the continued exponential, we plot $C_v(K)/C_v(0.06)$ around the critical points $1/K_c$ for $d=3$ sc, bcc and fcc in Fig. 10(a). The $C_v(K)$ is normalised with an arbitrarily high value $C_v(0.06)$, and it can be observed that this captures a similar singular nature for sc, bcc and fcc in the vicinity of $1/K_c$ from the low-temperature side. The critical values for $1/K_c$ in the literature \cite{Baker1963} are given by $1/0.4407=2.2692$, $1/0.2217=4.5102$, $1/0.1575=6.3505$, $1/0.1021=9.7923$ for sq, sc, bcc and fcc correspondingly. Similarly, $C_v(K)$ for $d=2$ sq seems to possess singular nature from the high-temperature side in Fig. 10(b). This unique behaviour may be related to the Kramers-Wannier duality on square lattice \cite{PhysRevST} where the strong coupling at low temperature gets mapped to the weak coupling at high temperature and vice-versa. From these curves, it is deduced that the value for exponent $\alpha$ at their corresponding $K_c$ is $\alpha=0.1026, 0.1193$ for three-dimensional bcc, fcc and $\alpha=-0.0138$ for two-dimensional sq respectively. These seem to be comparable with values $\alpha=0.11$ for $d=3$ and $\alpha=0$ for $d=2$ Ising models \cite{kardar2007statistical}.
  
\subsubsection{Migdall-Kadanoff position space renormalization}
Kadanoff's renormalization scheme was used on two-dimensional Ising models using successive approximations to control the divergent long-range interactions, and the correlation length critical exponent $\nu_{Ising} \approx 1$ was extracted  \cite{Kadanoff:1976jb,MARTINELLI1981201,CARACCIOLO1981405}. However, the primitive renormalization approach \cite{Kadanoff:1976jb} does not produce a reliable estimate of $\nu_{Ising}$, which was systematically improved later \cite{MARTINELLI1981201,CARACCIOLO1981405}. Similarly, we take the most straightforward position space renormalization scheme of the one-dimensional Ising model, where the decimation of every alternate spin on the lattice essentially reduces the $N'$ degrees of freedom by a rescaling factor of $b=2$ in Eq.(\ref{Ising}) \cite{kardar2007statistical}. Then we introduce new interactions in the renormalization scheme to account for long-range behaviour and implement continued exponential to approximate the divergent interactions controlled by a free parameter $a$. Further, Migdal-Kadanoff bond moving approximation is used to obtain exponent $\nu_{Ising}$ for phase transitions on fractal systems with non-integer dimensions $1<d<2$. 

There is a simple mapping between the original spins to the renormalized spins ($\{\sigma_i\} \mapsto \{\sigma'_i\}$) in the partition function $Z$ (Eq. \ref{Ising}) after summing over the decimated spins {$s_i$} as \cite{kardar2007statistical}
\begin{multline}
     \sum_{\{\sigma'_i\}}^{N'/2} \sum_{\{s_i\}}^{N'/2} \exp{ \left[ \sum_{i=1}^{N'/2} B(\sigma'_i,s_i)+B(s_i,\sigma'_{i+1})\right]} = \\ \sum_{\{\sigma'_i\}}^{N'/2} \prod_{i=1}^{N'/2}  \left[ \sum_{s_i=\pm1} \hbox{e}^{ B(\sigma'_i,s_i)+B(s_i,\sigma'_{i+1})}\right] \equiv \sum_{\{\sigma'_i\}}^{N'/2} \hbox{e}^{ \left[ \sum_{<i,i+1>}^{N'/2} B'(\sigma'_i,\sigma'_{i+1})\right]}  
     .
\end{multline}
Where the bond energy of the renormalized spins is
\begin{equation}
    B'(\sigma'_1,\sigma'_{2}) = \frac{h'}{2}(\sigma'_1+\sigma'_2) + K'\sigma'_1 \sigma'_2.
\end{equation}
The renormalized interactions are obtained from the assumption that the renormalized parameters ($h', K'$) are functions of ($h, K$) having similar formalism. We assume here that the renormalized parameters ($h', K'$) are analytical functions of ($h, K$) since it reflects Kadanoff's scaling idea \cite{KADANOFF:1967zz}. $h$ and $K$ are independent parameters that govern the continuous phase transition in a ferromagnetic-paramagnetic system around the point of criticality. The renormalized Hamiltonian per bond with the renormalized interactions is
\begin{multline}
   R(\sigma'_1,\sigma'_2) \equiv  \exp \left[ K'\sigma'_1 \sigma'_2 + \frac{h'}{2}(\sigma'_1 + \sigma'_2)\right] \\  = \sum_{s_1=\pm1} \exp \Bigg[ k_1 (K\sigma'_1 s_1 + K\sigma'_2 s_1 + K_2(\sigma'_1 s_1)(\sigma'_2 s_1)) +  h_1 \left(\frac{h}{2}(\sigma'_1 + s_1) + \frac{h}{2}(\sigma'_2 + s_1)\right) + \\ k_2 ((K\sigma'_1 s_1)^2 + (K\sigma'_2 s_1)^2 + K_2^2(\sigma'_1 s_1)^2(\sigma'_2 s_1)^2)+ \\ h_2 \left( \left(\frac{h}{2}\sigma'_1\right)^2 + \left(\frac{h}{2}\sigma'_2\right)^2 + 2\left(\frac{h}{2}s_1\right)^2 \right)  
   + \\ k_3 ((K\sigma'_1 s_1)^3 + (K\sigma'_2 s_1)^3 + K_2^3(\sigma'_1 s_1)^3(\sigma'_2 s_1)^3) + \\ h_3 \left( \left(\frac{h}{2}\sigma'_1\right)^3 + \left(\frac{h}{2}\sigma'_2\right)^3 + 2\left(\frac{h}{2}s_1\right)^3 \right) + \cdots   \Bigg],
\end{multline}
where $\{k_i\}$ and $\{h_i\}$ are the coefficients associated with the power series. The above expression $R(\sigma'_1,\sigma'_2)$ is the most generalized series expansion. Though it reduces the degrees of freedom, the renormalisation procedure typically introduces more interactions than in the original Hamiltonian. We introduce a new long-range interaction due to the renormalization procedure between pair of spin pairs $(\sigma'_1 s_1)$ and $(\sigma'_2 s_1)$ with a probabilistic weight $K_2$ in the probabilistic description. Since ${\sigma'}_i^2, s_1^2=1$ and ${\sigma'}_i^{3}=\sigma'_i, s_1^3=s_1$, the probabilistic weight is not going to change due to them. So we do not include the power series terms of the individual spins and spin pairs for determining the renormalized interactions. Rather, we include only the terms with $K_2$ assigning the weight $K_2=K^a$. We assign $a$ as the parameter that controls the strength of long-range interactions. To confine the long-range interactions, we convert the analytical function to a continued exponential such that $k_i = (i+1)^{i-1}/i!$ and renormalized interactions $R(\sigma'_1,\sigma'_2)$ becomes
\begin{multline}
    R(\sigma'_1,\sigma'_2) = \exp \left[ K'\sigma'_1 \sigma'_2 + \frac{h'}{2}(\sigma'_1 + \sigma'_2)\right] = \\  \hbox{e}^{K^a\sigma'_1 \sigma'_2 \hbox{e}^{K^a\sigma'_1 \sigma'_2\hbox{e}^{K^a\sigma'_1 \sigma'_2\hbox{e}^{\cdots}}}} \sum_{s_1=\pm1} \exp \left[K\sigma'_1 s_1 + K\sigma'_2 s_1+\frac{h}{2}(\sigma'_1 + \sigma'_2)+h s_1\right].
\end{multline}
All the possible configurations of the spins are considered,
\begin{fleqn}
\begin{multline*}
 R(+1,+1) = \hbox{e}^{K'} \hbox{e}^{h'} = \hbox{e}^{K^a \hbox{e}^{K^a\hbox{e}^{K^a\hbox{e}^{\cdots}}}} \hbox{e}^{h} \big( \hbox{e}^{2K} \hbox{e}^{h} + \hbox{e}^{-2K} 
    \hbox{e}^{-h} \big), \\
R(-1,-1) = \hbox{e}^{K'} \hbox{e}^{-h'} = \hbox{e}^{K^a \hbox{e}^{K^a\hbox{e}^{K^a\hbox{e}^{\cdots}}}} \hbox{e}^{-h} \big( \hbox{e}^{-2K} \hbox{e}^{h} + \hbox{e}^{2K} 
    \hbox{e}^{-h} \big), \\
R(+1,-1) = \hbox{e}^{-K'} = \hbox{e}^{-K^a \hbox{e}^{-K^a\hbox{e}^{-K^a\hbox{e}^{\cdots}}}} \big( \hbox{e}^{h}+\hbox{e}^{-h} \big), \\ 
R(-1,+1) = \hbox{e}^{-K'} = \hbox{e}^{-K^a \hbox{e}^{-K^a\hbox{e}^{-K^a\hbox{e}^{\cdots}}}} \big( \hbox{e}^{h}+\hbox{e}^{-h}  \big). \\
    \end{multline*}
    \end{fleqn}
The solutions for the above take the form of
 \begin{equation}
      \hbox{e}^{4K'} = \frac{\hbox{e}^{2K^a \hbox{e}^{K^a\hbox{e}^{K^a\hbox{e}^{\cdots}}}}\big( \hbox{e}^{2K} \hbox{e}^{h} + \hbox{e}^{-2K} 
    \hbox{e}^{-h} \big)\big( \hbox{e}^{-2K} \hbox{e}^{h} + \hbox{e}^{2K} 
    \hbox{e}^{-h} \big)}{\hbox{e}^{-2K^a \hbox{e}^{-K^a\hbox{e}^{-K^a\hbox{e}^{\cdots}}}}\big( \hbox{e}^{h}+\hbox{e}^{-h}  \big)^2}, \label{KK'}
 \end{equation}
 \begin{equation}
     \hbox{e}^{2h'} = \frac{\hbox{e}^{2h}\big( \hbox{e}^{2K} \hbox{e}^{h} + \hbox{e}^{-2K} 
    \hbox{e}^{-h} \big)}{ \big( \hbox{e}^{-2K} \hbox{e}^{h} + \hbox{e}^{2K} 
    \hbox{e}^{-h} \big)}.
 \end{equation}
 The critical point $K_c$ can be found by setting $h=0$ since its a symmetry-breaking term, so studying the characteristics of $K$ in a subspace where the symmetry is maintained such that $h=0$ implies $h'=0$,
 \begin{equation}
     \hbox{e}^{2K'} = \frac{\hbox{e}^{K^a \hbox{e}^{K^a\hbox{e}^{K^a\hbox{e}^{\cdots}}}}\big( \hbox{e}^{-2K} + \hbox{e}^{2K}  \big)}{2\hbox{e}^{-K^a \hbox{e}^{-K^a\hbox{e}^{-K^a\hbox{e}^{\cdots}}}}}.
 \end{equation}
 We obtain the fixed point for the above recursion relation of $K$ such that the sequence 
 \begin{equation}
     \frac{1}{2}\hbox{ln}\left(\frac{\hbox{e}^{K_c^a}(\hbox{e}^{-2K_c}+\hbox{e}^{2K_c})}{2\hbox{e}^{-K_c^a}}\right),  \frac{1}{2}\hbox{ln}\left(\frac{\hbox{e}^{K_c^a\hbox{e}^{K_c^a}}(\hbox{e}^{-2K_c}+\hbox{e}^{2K_c})}{2\hbox{e}^{-K_c^a\hbox{e}^{-K_c^a}}}\right),\frac{1}{2}\hbox{ln}\left(\frac{\hbox{e}^{K_c^a\hbox{e}^{K_c^a\hbox{e}^{K_c^a}}}(\hbox{e}^{-2K_c}+\hbox{e}^{2K_c})}{2\hbox{e}^{-K_c^a\hbox{e}^{-K_c^a\hbox{e}^{-K_c^a}}}}\right),\cdots
\end{equation}
approaches a converged value of $K_c$, the point of criticality where we are interested in studying the interactions. For $d>1$ where the ordering happens, we use the Migdall-Kadanoff bond moving approximation where the bond strengthens by a factor $2^{d-1}$ ($K$ becomes $2^{d-1}K$) for any $d$-dimensional hypercubic lattice in Eq.(\ref{KK'}). Using this formalism, we get $K_c=0.1271$ for $d=2$ taking $a=2$. The position of $K_c$ might be wrong compared to the two-dimensional Ising model ($K_c=0.4407$), but the behaviour of the correlation length, which can be described by critical exponent $\nu_{Ising}$ obtained from the expression below remains unchanged,
\begin{equation}
b^{\frac{1}{\nu_{Ising}}} = \frac{\partial K'}{\partial K} \Bigr|_{\substack{K_c}}.
\end{equation}
The wrong $K_c$ values can be attributed due to arbitrary $\{k_i\}$ values taken in $R(\sigma'_1,\sigma'_2)$ for easy evaluation. However, using the above expression, we can obtain $\nu_{Ising}=1.019$ for $d=2$, which is comparable with the exact value $\nu_{Ising}=1$ \cite{onsager}. We also obtain that as $d\rightarrow\infty$, $\nu_{Ising}\rightarrow$1  for $a=2$ which seems to be the typical behaviour of Migdall-Kadanoff bond moving approximation on $d$-dimensional hypercubic lattice \cite{kardar2007statistical}.
We numerically deduce the strength of the long-range interactions $a\propto d$ dimensionality of the Ising model for describing the phase transitions on fractal systems. For $1<d<2$, we derive the empirical relation $a\sim 0.9d^{1.082}$, which can predict the behaviour of the correlation length at a critical point similar to that of the previous studies of Pad\'e-Borel summation on Callan–Symanzik scheme of renormalization \cite{Holovatch1993}. Using this relation, we compute $\nu_{Ising}$ for non-integer dimensions and compare it with literature in Table \rom{1}.   \begin{table}[ht]
\small
\begin{center}
\caption{Critical exponent of $O(1)$ class $\nu_{Ising}$ for $1<d<2$.}

\begin{tabular}{|c|c|c|c|c|c|c|}

\hline  
 $d$ & 1.25 &  1.375 & 1.5 & 1.650 & 1.750 & 1.875 \\ 
 \hline
 $\nu_{Ising}$ & 2.9879 & 1.9414 & 1.5542 & 1.3162 & 1.2158 & 1.1253 \\
 \hline
  $\nu_{Ising}$ \cite{Holovatch1993} & 2.593 & 1.983 & 1.627 & 1.353 & 1.223 & 1.098 \\
  \hline
\end{tabular}

\label{table 16}
\end{center}
\end{table} 
\section{Critical exponents of quantum phase transitions \\ (Gross-Neveu-Yukawa models)}
Recently Dirac materials \cite{reviewdm} and Weyl semimetals \cite{reviewws} have been interesting to study since they are believed to undergo second-order quantum phase transitions under particular scenarios. Such transitions are experimentally yet to be verified; however, their realization is theorised and studied with relevant order parameters related to the systems \cite{DW1,DW2,DW3,DW4,DW5,DW6,DW7,DW8}. However, since quantum phase transitions can happen at $T=0$, reduced temperature $|T-T_c|$ in the field-theoretic description is replaced with similar measures, such as variation of coupling constant from their critical values. While purely bosonic field theories describe $O(n)$ universality classes, a new class of universality emerges for Dirac and Weyl systems in the presence of fermionic fields described by Gross-Neveu-Yukawa (GNY) models \cite{ZINNJUSTIN1991,ROSENSTEIN1993}. Recently four-loop RG functions have been solved for different GNY models to address the critical exponents of such universality classes \cite{4loopgny}. However, they employed only simple diagonal Pad\'e approximants to evaluate the critical exponents as spurious poles riddled the other non-diagonal terms. Typically a thorough analysis is required when using Pad\'e-based methods with critical inspection for poles and their removal, as performed for different $\epsilon$-expansions in recent work \cite{KOMPANIETS2020}. We use the four-loop $\epsilon$ expansions \cite{4loopgny} to determine critical exponents related to these models implementing continued functions.
\subsection{Chiral Ising universality class}
The Chiral Ising model that can describe quantum phase transitions is a modification of the field-theoretic Ising model where fermions ($\psi$) are coupled to a scalar field ($\phi$) with Yukawa coupling. There are additional critical exponents in these GNY models associated with the RG gamma functions of the real scalar field and fermions, anomalous dimensions of bosons ($\eta_{\phi}$) and fermions ($\eta_{\psi}$). The difference in the description of such GNY models has been generalized by a parameter $N$, the number of fermion flavours of the four-component Dirac fermion in the model. These models have a range of applicability depending on $N$. The most physically relevant systems are the semimetal-charge density wave transition of electrons in graphene for $N=2$ \cite{PRL-CDW} and semimetal-insulator transition of spinless fermions on honeycomb lattice for $N=1$. These systems have also been studied using other methods such as non-perturbative functional renormalization group (FRG) \cite{FRG-ISING1,FRG-ISING2}, quantum Monte-Carlo simulations (QMC) \cite{QMC-ISING1,QMC-ISING2,QMC-ISING3} and CB \cite{CB-ISING1,CB-ISING2} to calculate their corresponding critical exponents. For $N=1/4$, this model is theorised to exhibit emergent supersymmetry properties on the boundary of topological superconductors \cite{SUSY}. \\ We implement continued exponential fraction, continued exponential and continued exponential with Borel-Leroy transformation (Eq.s (2), (3) and (4)) to obtain estimates for exponents \cite{4loopgny} \begin{subequations}
    \begin{align}
        1/\nu &=2 - 0.9524 \epsilon +0.007225 \epsilon^2 - 0.09487 \epsilon^3 - 0.01265 \epsilon^4\,\\
         \eta_{\phi} &= 0.5714\epsilon + 0.1236\epsilon^2 - 0.02789\epsilon^3 + 0.1491\epsilon^4,\\
         \eta_{\psi} &=0.07143\epsilon - 0.006708\epsilon^2 - 0.02434\epsilon^3+ 0.01758\epsilon^4, \\
         \omega& =\epsilon - 0.3525 \epsilon^2 + 0.4857 \epsilon^3 - 1.338 \epsilon^4,
    \end{align}
\end{subequations} for $N=2$ in $d=2+1$. These estimates at consecutive orders for $1/\nu$, $\eta_{\phi}$, $\eta_{\psi}$ and $\omega$ are illustrated, compared with QMC predictions \cite{QMC-ISING1} in Fig.s 11(a), 11(b), 12(a) and 12(b), respectively. 
\begin{figure}[!htp]
\centering
\begin{subfigure}{0.495\textwidth}
\includegraphics[width=1\linewidth, height=6cm]{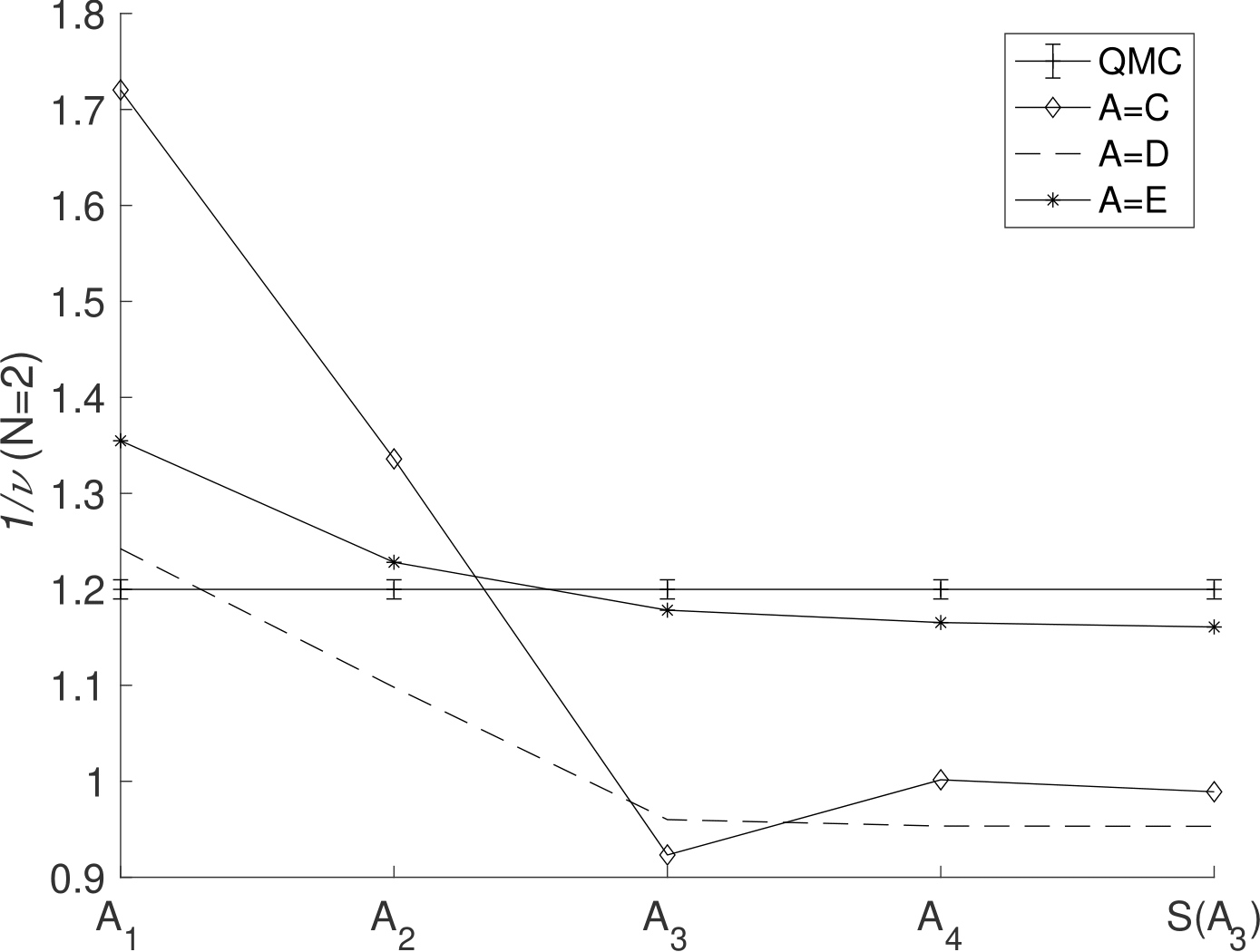} 
\caption{Estimates of $1/\nu$ at successive orders.}

\end{subfigure}
\begin{subfigure}{0.495\textwidth}
\includegraphics[width=1\linewidth, height=6cm]{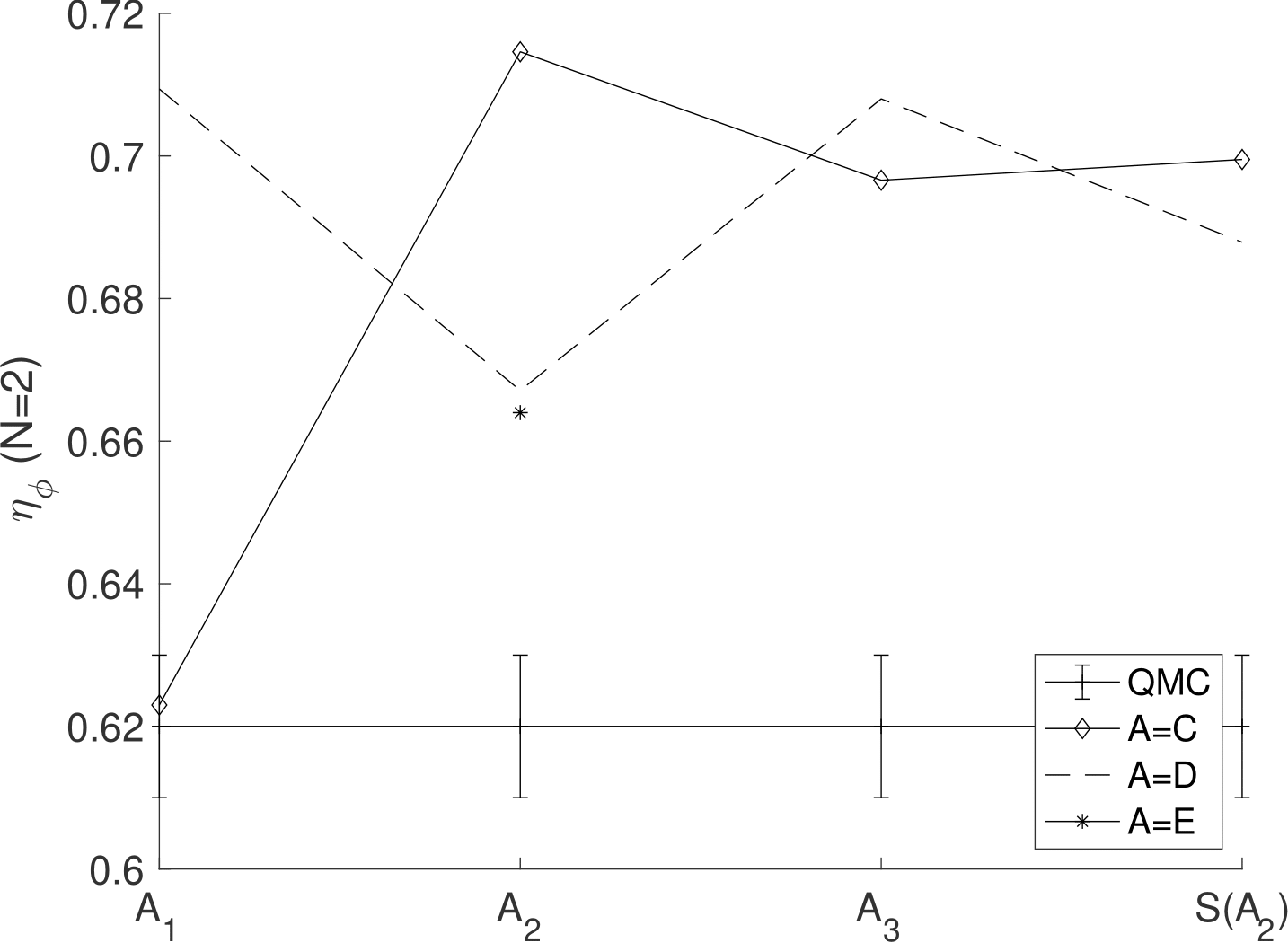}
\caption{Estimates of $\eta_{\phi}$ at successive orders.}

\end{subfigure}

\caption{Comparing Chiral Ising universality class $1/\nu$ and $\eta_{\phi}$ of $N=2$  with QMC results \cite{QMC-ISING1}.}
\end{figure} 
\begin{figure}[!htp]
\centering
\begin{subfigure}{0.495\textwidth}
\includegraphics[width=1\linewidth, height=6cm]{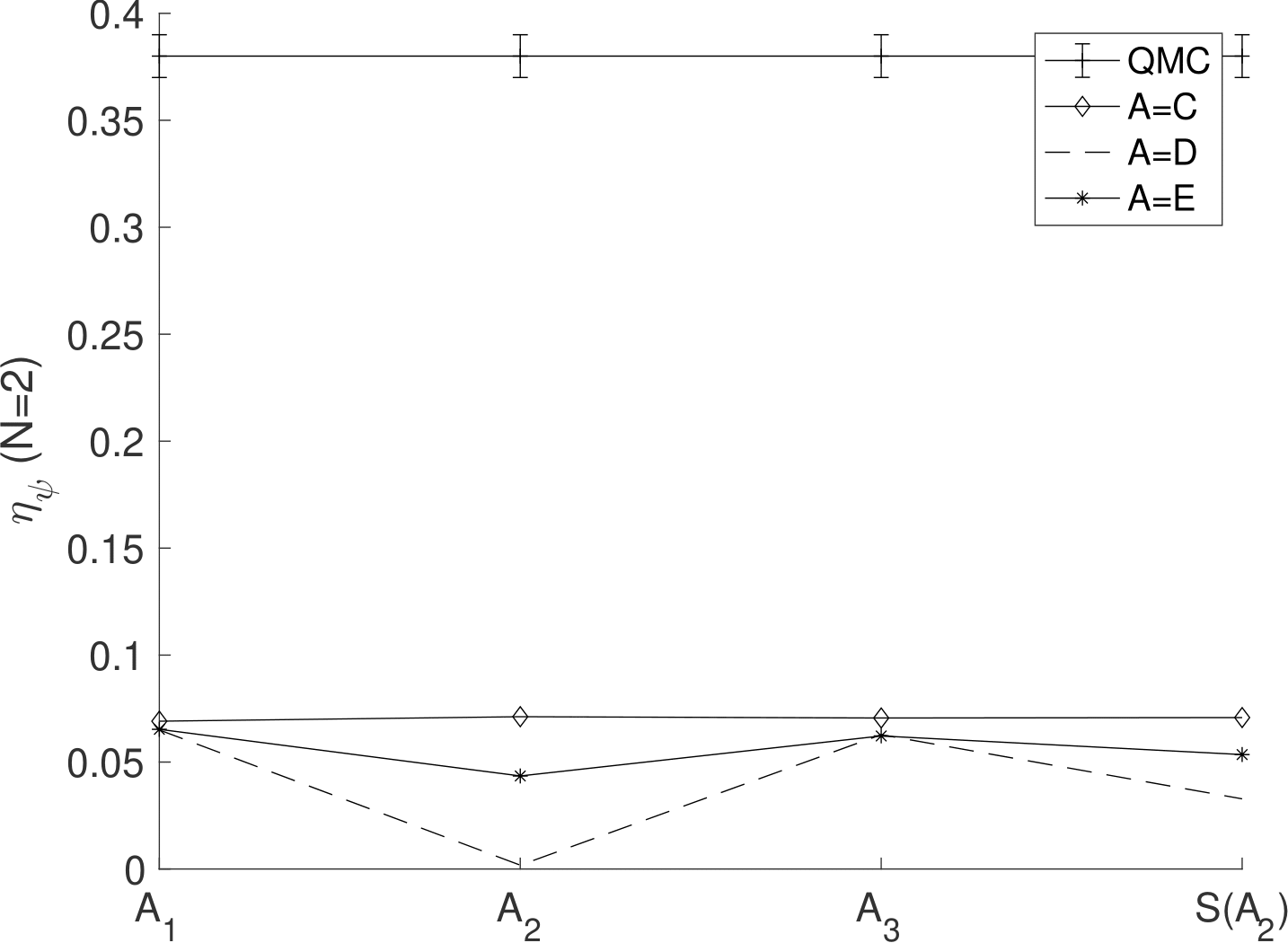} 
\caption{Estimates of $\eta_{\psi}$ at successive orders compared with QMC.}

\end{subfigure}
\begin{subfigure}{0.495\textwidth}
\includegraphics[width=1\linewidth, height=6cm]{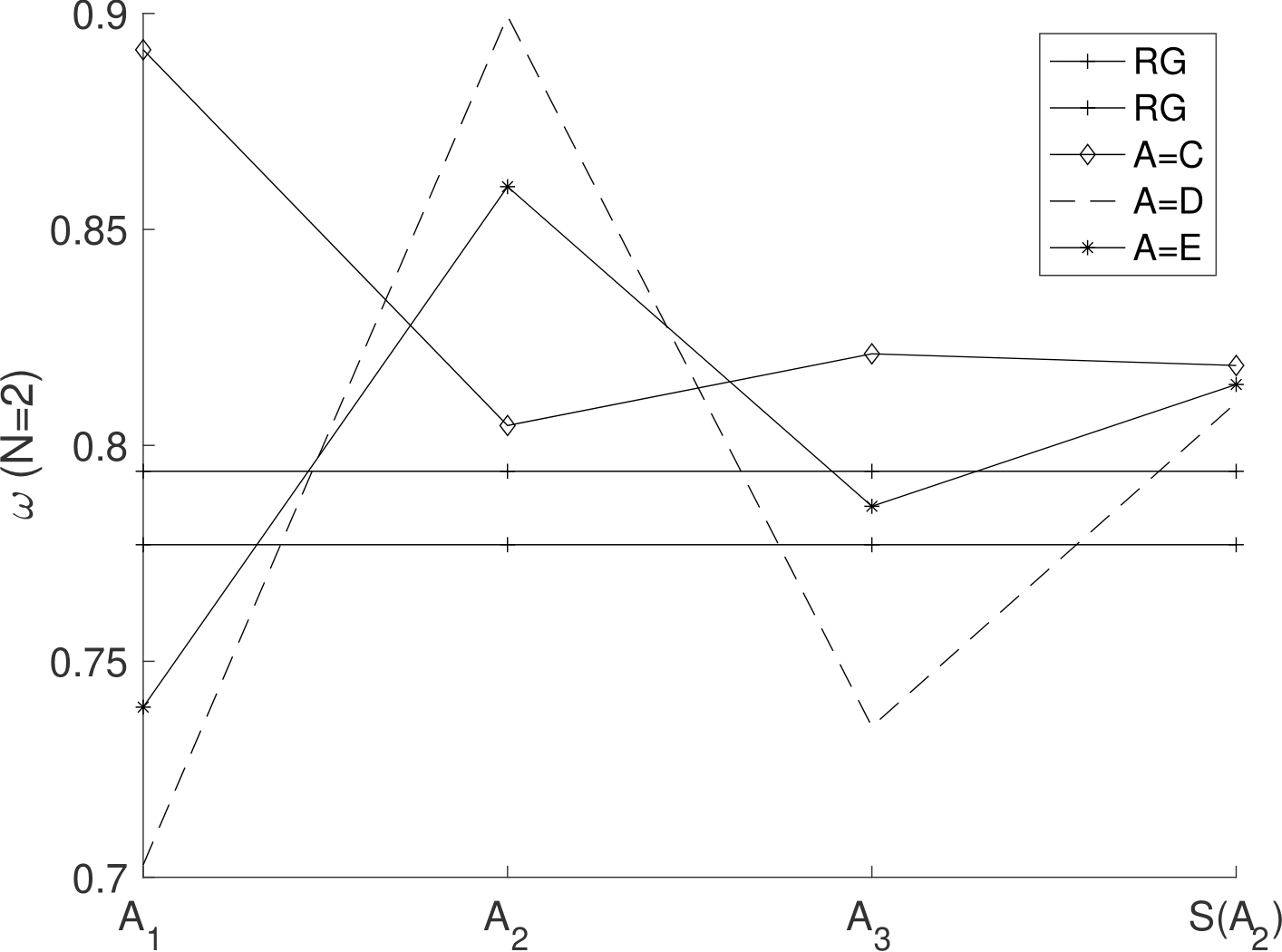}
\caption{Estimates of $\omega$ at successive orders compared with RG.}

\end{subfigure}

\caption{Comparing Chiral Ising universality class $\eta_{\psi}$ and $\omega$ of $N=2$  with QMC \cite{QMC-ISING1} and RG \cite{4loopgny} estimates.}
\end{figure} 
\\Similarly, we obtain estimates for exponents \cite{4loopgny} \begin{subequations}
    \begin{align}
        1/\nu &=2 - 0.8347 \epsilon -0.0057 \epsilon^2 - 0.0603 \epsilon^3 - 0.0903 \epsilon^4,\\
         \eta_{\phi} &= 0.4\epsilon + 0.1025\epsilon^2 - 0.0632\epsilon^3 + 0.1986\epsilon^4,\\
         \eta_{\psi} &=0.1\epsilon - 0.0102\epsilon^2 - 0.0330\epsilon^3+ 0.0507\epsilon^4, 
    \end{align}
\end{subequations} for $N=1$ in $d=2+1$. These estimates at consecutive orders for $1/\nu$, $\eta_{\phi}$ and $\eta_{\psi}$ are illustrated, compared with predictions from QMC \cite{QMC-ISING2,QMC-ISING3}, CB \cite{CB-ISING2} predictions in Fig.s 13(a), 13(b) and 14(a), respectively. \begin{figure}[!htp]
\centering
\begin{subfigure}{0.495\textwidth}
\includegraphics[width=1\linewidth, height=6cm]{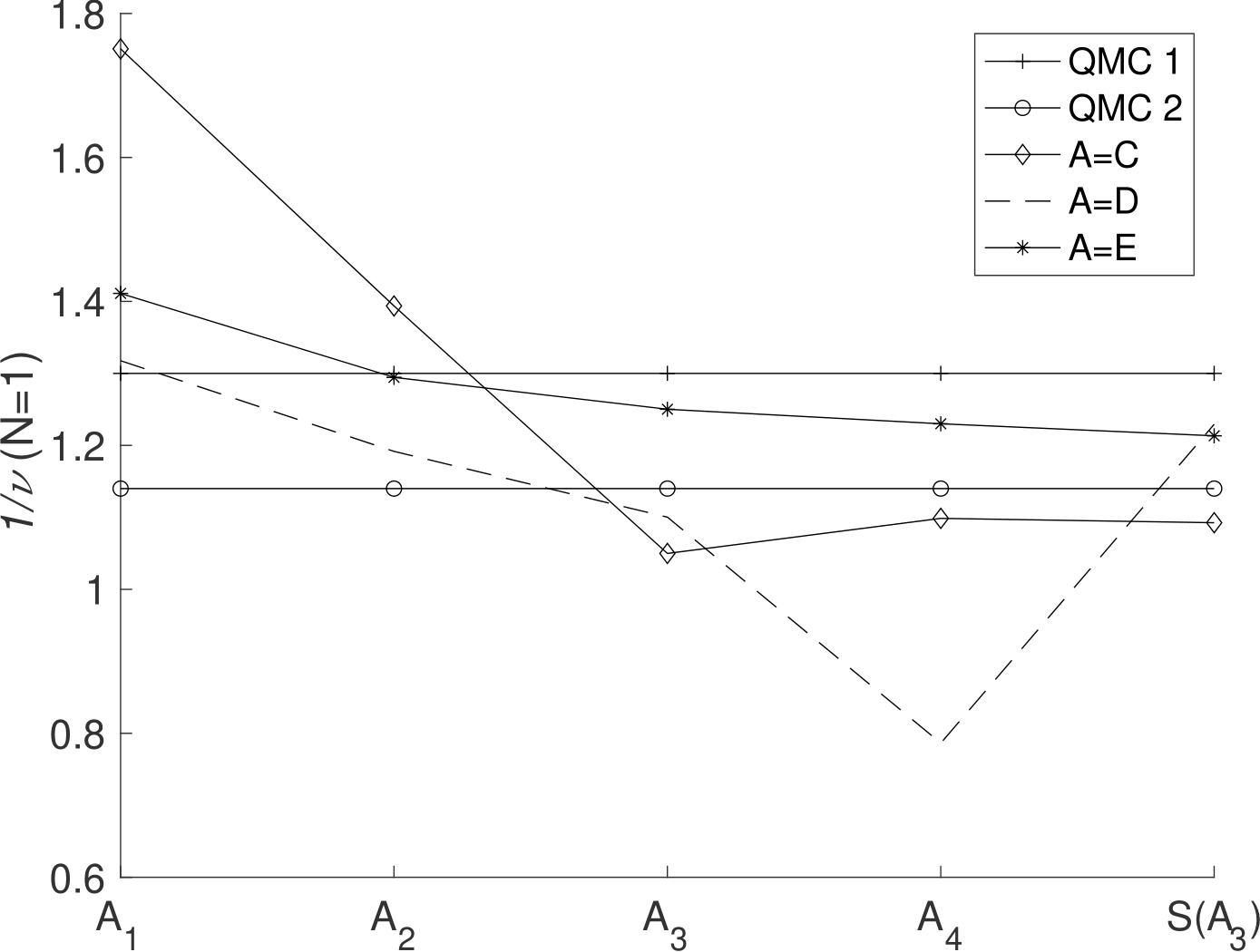} 
\caption{Estimates of $1/\nu$ at successive orders.}

\end{subfigure}
\begin{subfigure}{0.495\textwidth}
\includegraphics[width=1\linewidth, height=6cm]{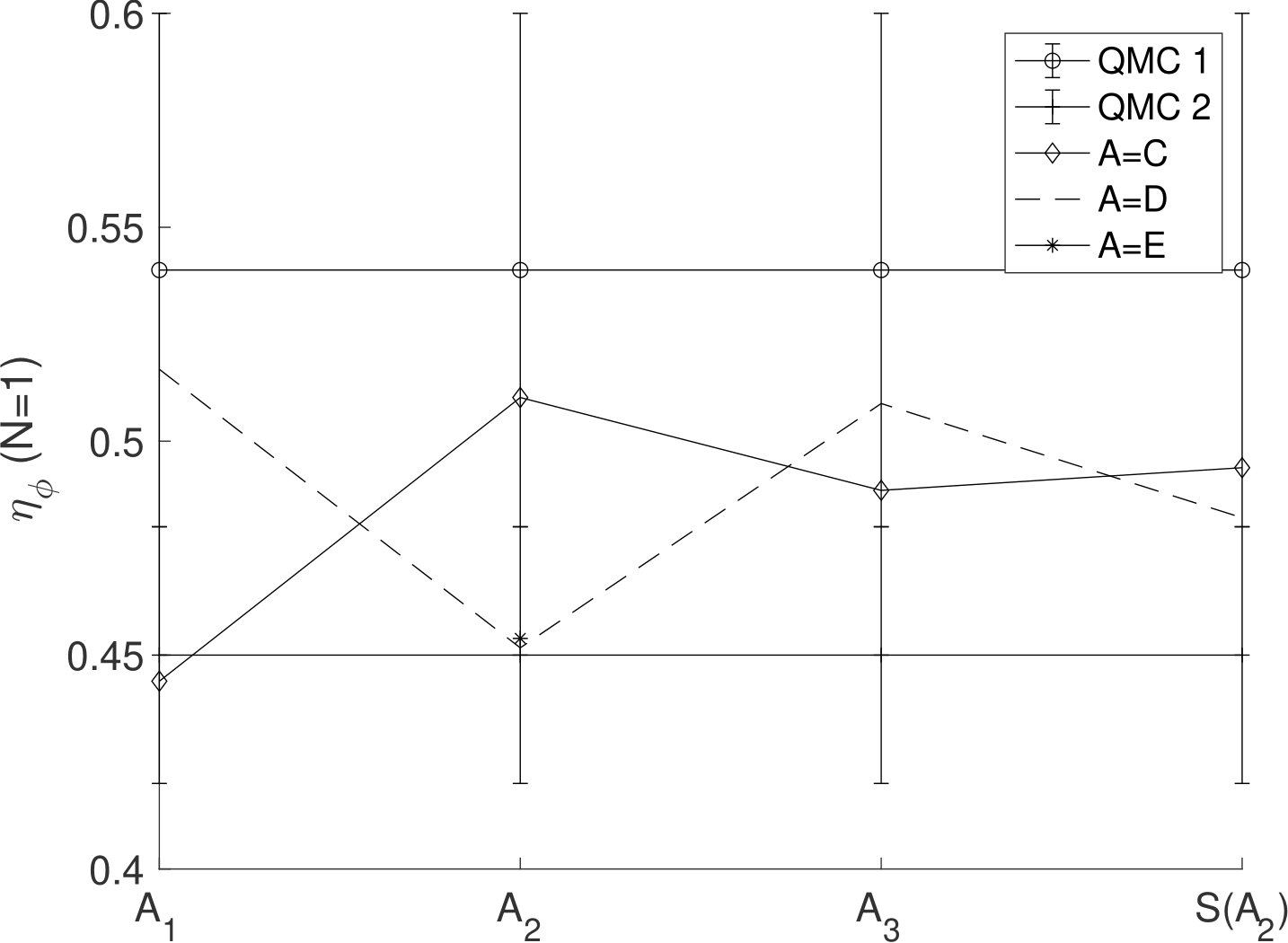}
\caption{Estimates of $\eta_{\phi}$ at successive orders.}

\end{subfigure}

\caption{Comparing Chiral Ising universality class $1/\nu$ and $\eta_{\phi}$ of $N=1$  with QMC \cite{QMC-ISING2,QMC-ISING3} estimates.}
\end{figure} 
\begin{figure}[!htp]
\centering
\begin{subfigure}{0.495\textwidth}
\includegraphics[width=1\linewidth, height=6cm]{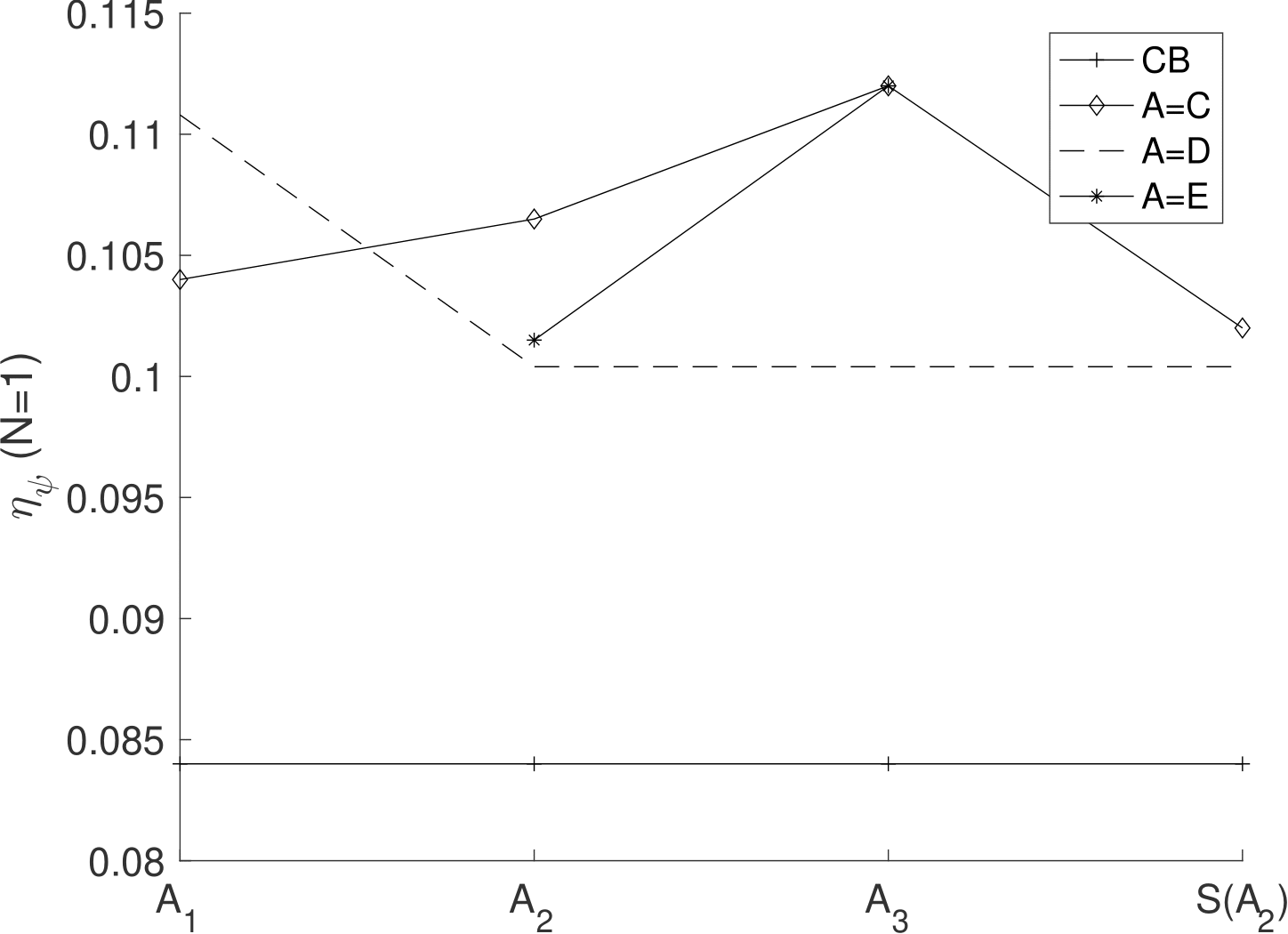} 
\caption{Estimates of $\eta_{\psi}$ ($N=1$) at successive orders compared with CB.}

\end{subfigure}
\begin{subfigure}{0.495\textwidth}
\includegraphics[width=1\linewidth, height=6cm]{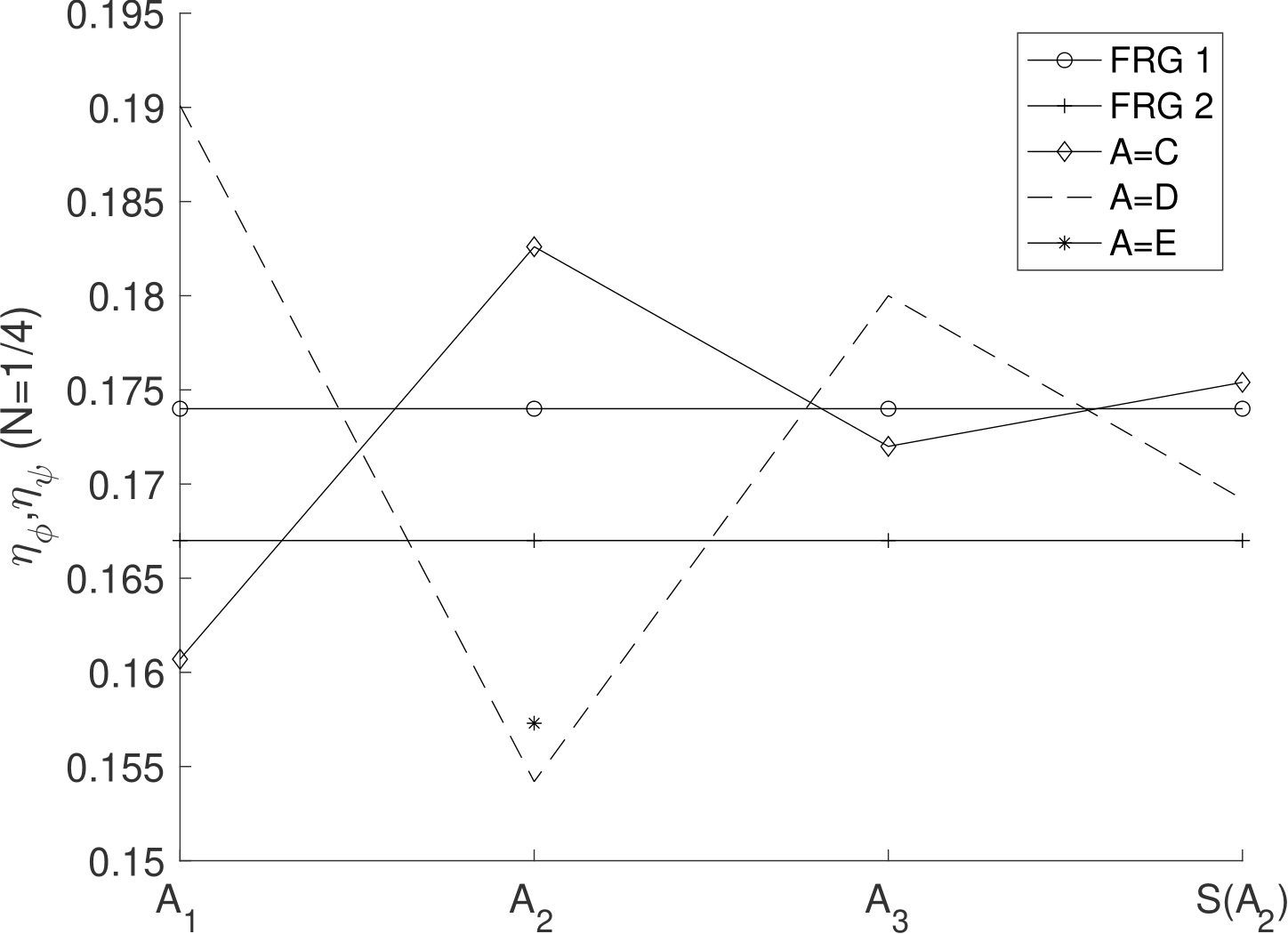}
\caption{Estimates of $\eta_{\phi},\eta_{\psi}$ ($N=1/4$) at successive orders compared with FRG.}

\end{subfigure}

\caption{Comparing Chiral Ising universality class $\eta_{\psi}$ of $N=1,1/4$ and $\eta_{\phi}$ of $N=1/4$  with CB \cite{CB-ISING2} and FRG \cite{Gies2017} estimates.}
\end{figure} \\ And, similarly we obtain estimates for exponents \cite{4loopgny}  \begin{subequations}
    \begin{align}
        1/\nu &=2 - 0.5714 \epsilon -0.0204 \epsilon^2 + 0.0240 \epsilon^3 - 0.0596 \epsilon^4,\\
         \eta_{\phi} &= \eta_{\psi}= 0.1429\epsilon + 0.0408\epsilon^2 - 0.0480\epsilon^3 + 0.1193\epsilon^4,\\
         \omega& =\epsilon - 0.4286\epsilon^2 + 1.1763 \epsilon^3 - 4.0099 \epsilon^4,
    \end{align}
\end{subequations} for $N=1/4$ in $d=2+1$ which are illustrated, compared with FRG predictions \cite{Gies2017} in Fig.s 15(a), 14(b), 15(b), respectively. \begin{figure}[!htp]
\centering
\begin{subfigure}{0.495\textwidth}
\includegraphics[width=1\linewidth, height=6cm]{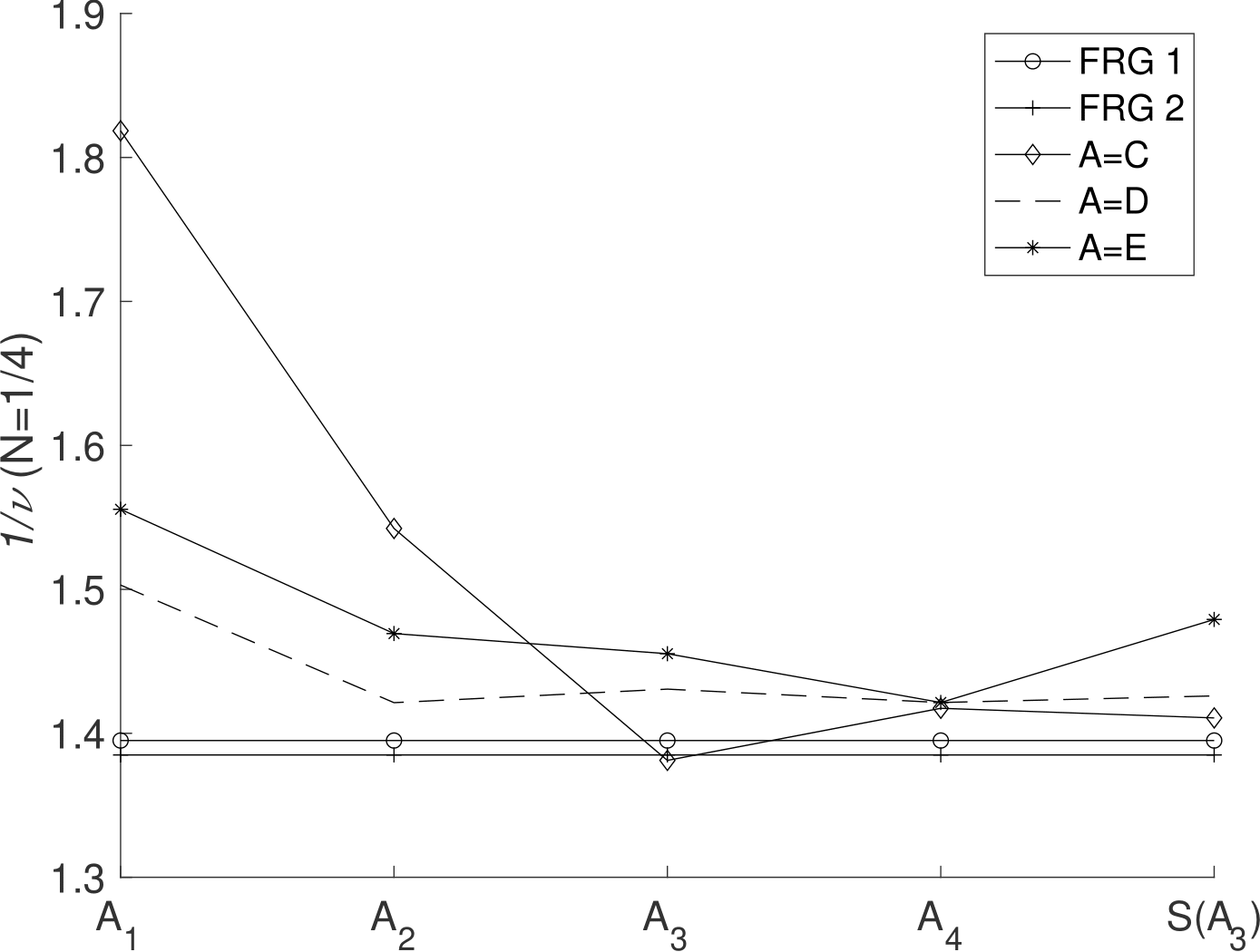} 
\caption{Estimates of $1/\nu$ at successive orders.}

\end{subfigure}
\begin{subfigure}{0.495\textwidth}
\includegraphics[width=1\linewidth, height=6cm]{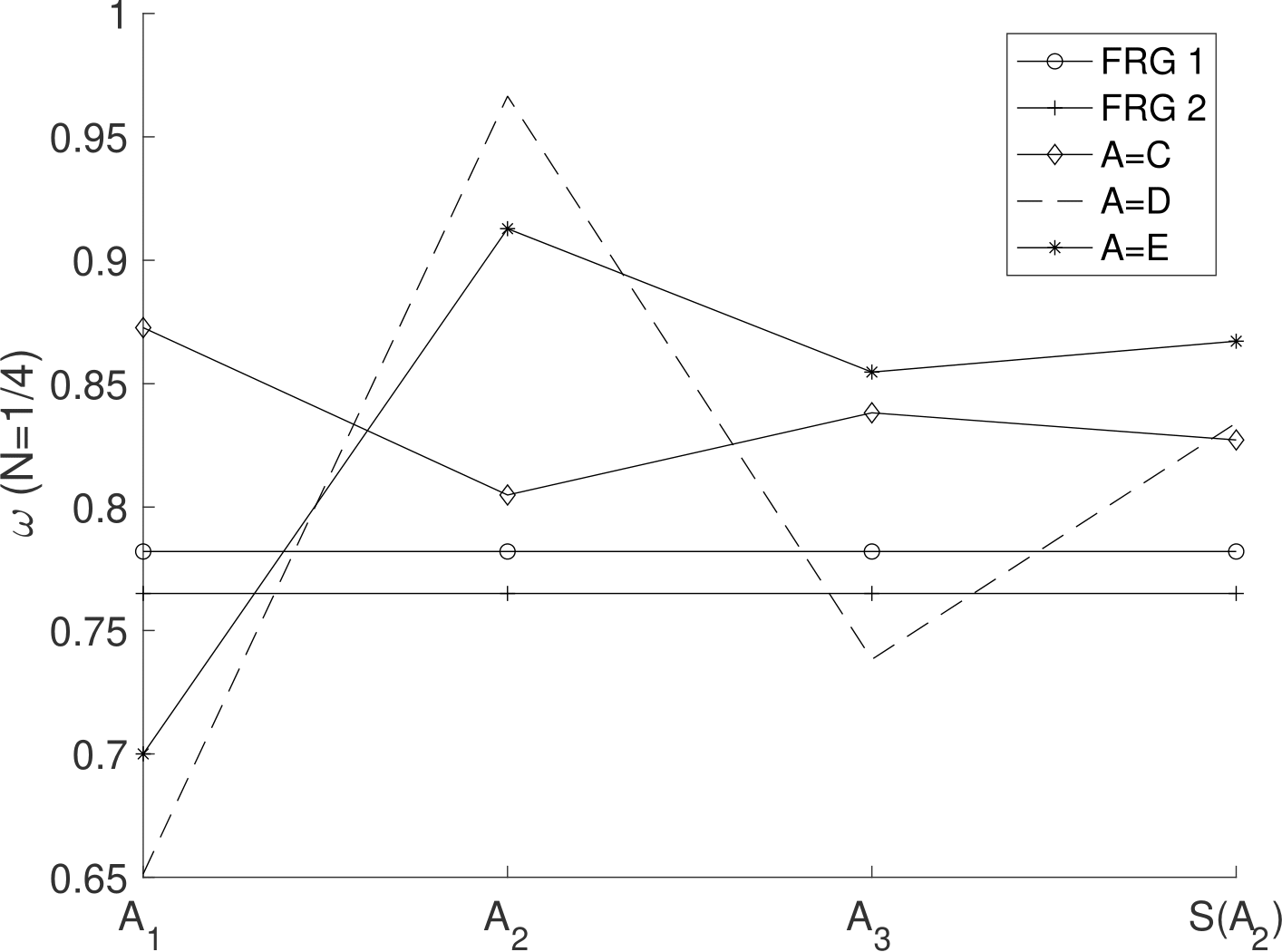}
\caption{Estimates of $\omega$ at successive orders.}

\end{subfigure}

\caption{Comparing Chiral Ising universality class $1/\nu$ and $\omega$ of $N=1/4$  with FRG \cite{Gies2017} estimates.}
\end{figure} We observe that these predictions tabulated in Table \rom{2} are mostly comparable with existing literature from FRG, QMC, and CB and are precisely compatible with Pad\'e resummation of RG \cite{4loopgny}. Estimates seem to undershoot or overshoot slightly, whereas there is large uncertainty in predicting anomalous fermion dimension $\eta_{\psi}$ for $N=2,1$ from different approaches. When handling $\eta_{\phi}$, $\eta_{\psi}$ with continued exponential with Borel-Leroy transformation, spurious poles were encountered, where estimates are not available. \begingroup \scriptsize
\setlength{\tabcolsep}{0.5pt} 
\renewcommand{\arraystretch}{1}
\begin{table}[htp!]
\small
\begin{center}
\caption{Critical exponents of Chiral Ising universality class $1/\nu$, $\eta_{\phi}$, $\eta_{\psi}$ and $\omega$ for $N=2,1,1/4$ . Our values derived from continued functions ($\{C,D,E\}$) are compared with recent literature.}

\begin{tabular}{ | c | c | c | c | c | }
\hline  
$N$ & $1/\nu$ & $\eta_{\phi}$ &  $\eta_{\psi}$ & $\omega$ \\ 
\hline
 2 & \begin{tabular}{c c c c c}
      &  0.989(45) ($S(C_3)$)  \\
      &  0.9531(35) ($S(D_3)$) \\
      &  1.1608(88) ($S(E_3)$) \\
      & 0.931, 0.945  \cite{4loopgny} \\
      & 0.994(2) \cite{FRG-ISING1} (FRG) \\
      & 1.20(1) \cite{QMC-ISING1} (QMC)
 \end{tabular} & \begin{tabular}{c c c c c}
      & 0.699(11) ($S(C_2)$) \\
      & 0.688(31)($S(D_2)$) \\
      & 0.664 ($E_2$)\\
      & 0.7079, 0.6906 \cite{4loopgny}\\
      & 0.742 \cite{CB-ISING2} (CB)\\
      & 0.7765 \cite{FRG-ISING1} (FRG) \\
      & 0.62(1) \cite{QMC-ISING1} (QMC)
 \end{tabular} & \begin{tabular}{c c c c c}
      & 0.0708 ($S(C_2)$) \\
      & 0.033(46) ($S(D_2)$) \\
      & 0.054(14) ($S(E_2)$) \\
      & 0.0539, 0.0506 \cite{4loopgny} \\
      & 0.044 \cite{CB-ISING2} (CB)\\
      & 0.0276 \cite{FRG-ISING1} (FRG) \\
      & 0.38(1) \cite{QMC-ISING1} (QMC)
 \end{tabular} & \begin{tabular}{c c c c c}
      & 0.8185(96) ($S(C_2)$) \\
      & 0.81(12) ($S(D_2)$) \\
      & 0.814(51) ($S(E_2)$) \\
      & 0.794, 0.777 \cite{4loopgny} \\
 \end{tabular} \\
 \hline 
 1 & \begin{tabular}{c c c c c}
      & 1.093(27) ($S(C_3)$) \\
      & 1.23(38) ($S(D_3)$) \\
      & 1.213(18) ($S(E_3)$) \\
      & 1.101 \cite{4loopgny} \\
      & 1.075(4) \cite{FRG-ISING1} (FRG) \\
      & 1.14 \cite{QMC-ISING2} (QMC) \\
      & 1.30 \cite{QMC-ISING3} (QMC)
 \end{tabular} & \begin{tabular}{c c c c c}
      & 0.494(14) ($S(C_2)$) \\
      & 0.482(42) ($S(D_2)$) \\
      & 0.4539 ($E_2$) \\
      & 0.4969, 0.4872 \cite{4loopgny}\\
      & 0.5506 \cite{FRG-ISING1} (FRG) \\
      & 0.544 \cite{CB-ISING2} (CB)\\
      & 0.54(6) \cite{QMC-ISING2} (QMC) \\
      & 0.45(3)\cite{QMC-ISING3} (QMC)
 \end{tabular} & \begin{tabular}{c c c c c}
      & 0.1019(77) ($S(C_2)$) \\
      & 0.1004 ($S(D_2)$) \\
      & 0.1011 ($E_3$) \\
      & 0.0976, 0.0972 \cite{4loopgny} \\
      & 0.0645 \cite{FRG-ISING1} (FRG)\\
      & 0.084 \cite{CB-ISING2} (CB)
 \end{tabular} & -\\
 \hline
 1/4 & \begin{tabular}{c c c c c}
      &  1.411(21) ($S(C_3)$)\\
      & 1.426(7) ($S(D_3)$) \\
      & 1.479(46) ($S(E_3)$)\\
      & 1.415  \cite{4loopgny} \\
      & 1.385, 1.395 \cite{Gies2017} (FRG) 
 \end{tabular} & \begin{tabular}{c c c c c}
      & 0.1754(71) ($S(C_2)$)\\
      &  0.169(18) ($S(D_2)$) \\
      & 0.1573 ($E_2$) \\
      & 0.171, 0.170 \cite{4loopgny}\\
      & 0.167,0.174 \cite{Gies2017} (FRG)\\
      & 0.164\cite{CB-ISING1} (CB)
 \end{tabular} & \begin{tabular}{c c c c c}
      & 0.1754(71) ($S(C_2)$)\\
      &  0.169(18) ($S(D_2)$) \\
      & 0.1573 ($E_2$) \\
      & 0.171, 0.170 \cite{4loopgny}\\
      & 0.167,0.174 \cite{Gies2017} (FRG)\\
      & 0.164\cite{CB-ISING1} (CB)
 \end{tabular} & \begin{tabular}{c c c c c}
      &  0.827(22) ($S(C_2)$) \\
      &  0.83(16) ($S(D_2)$) \\
      &  0.867(35) ($S(E_2)$) \\
      & 0.843, 0.838 \cite{4loopgny} \\
       & 0.765, 0.782 \cite{Gies2017} (FRG)\\
 \end{tabular} \\
 \hline
 
\end{tabular}

\label{table 17}
\end{center}
\end{table} \normalsize

\subsection{Chiral XY universality class}
In the chiral XY model, Dirac fermions undergo continuous $U(1)$ symmetry breaking described by a complex scalar field. The physically interesting systems in this model which can describe the quantum criticality of superconducting states in graphene are for $N=2$ \cite{CHIRALXY}. This is related to Kekul\'e transition on two-dimensional graphene structures \cite{prlxy,prbxy1,prbxy2}. Another interesting application of this model is in surface states of three-dimensional topological insulators where emergent supersymmetry is theorised for $N=1/2$ \cite{CHIRALXY,susyxy}. We obtain the estimates of critical exponents \cite{4loopgny} \begin{subequations}
    \begin{align}
        1/\nu &=2 - 1.2\epsilon + 0.1829 \epsilon^2 - 0.3515 \epsilon^3 + 0.5164 \epsilon^4, \\
         \eta_{\phi} &= 0.6667 \epsilon + 0.1211 \epsilon^2 - 0.005048 \epsilon^3 + 0.1938\epsilon^4,\\
         \eta_{\psi} &=0.1667 \epsilon - 0.02722\epsilon^2 -0.05507\epsilon^3 + 0.04202\epsilon^4, \\
         \omega& =\epsilon-0.3783\epsilon +0.6271\epsilon^3 -1.853\epsilon^4,
    \end{align}
\end{subequations} for $N=2$ in $d=2+1$.  These estimates at consecutive orders for $1/\nu$, $\eta_{\phi}$, $\eta_{\psi}$ and $\omega$ are illustrated, compared with predictions from QMC \cite{MC-XY}, FRG \cite{FRG-XY} in Fig.s 16(a), 16(b), 17(a) and 17(b), respectively. 
\begin{figure}[!htp]
\centering
\begin{subfigure}{0.495\textwidth}
\includegraphics[width=1\linewidth, height=6cm]{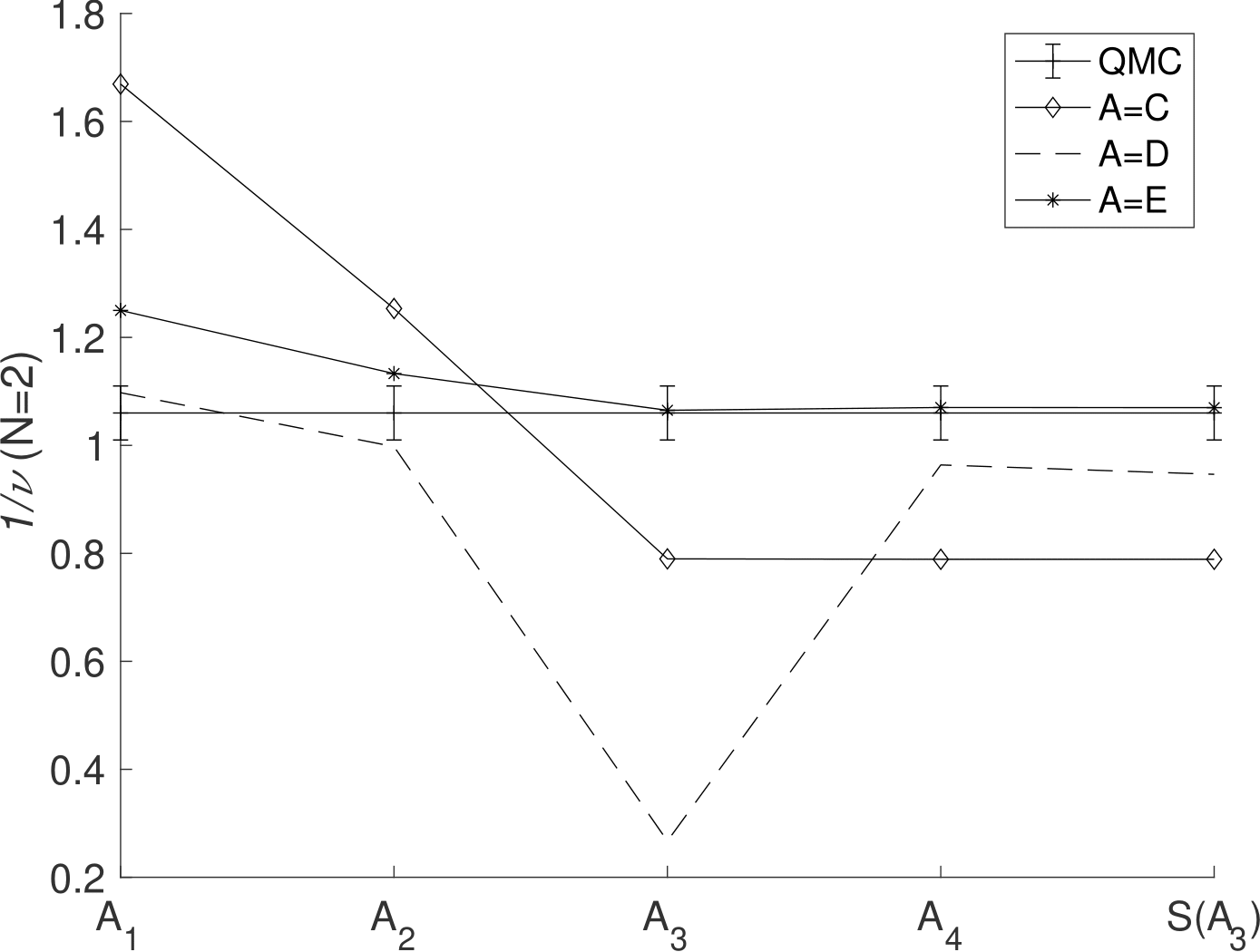} 
\caption{Estimates of $1/\nu$ at successive orders.}

\end{subfigure}
\begin{subfigure}{0.495\textwidth}
\includegraphics[width=1\linewidth, height=6cm]{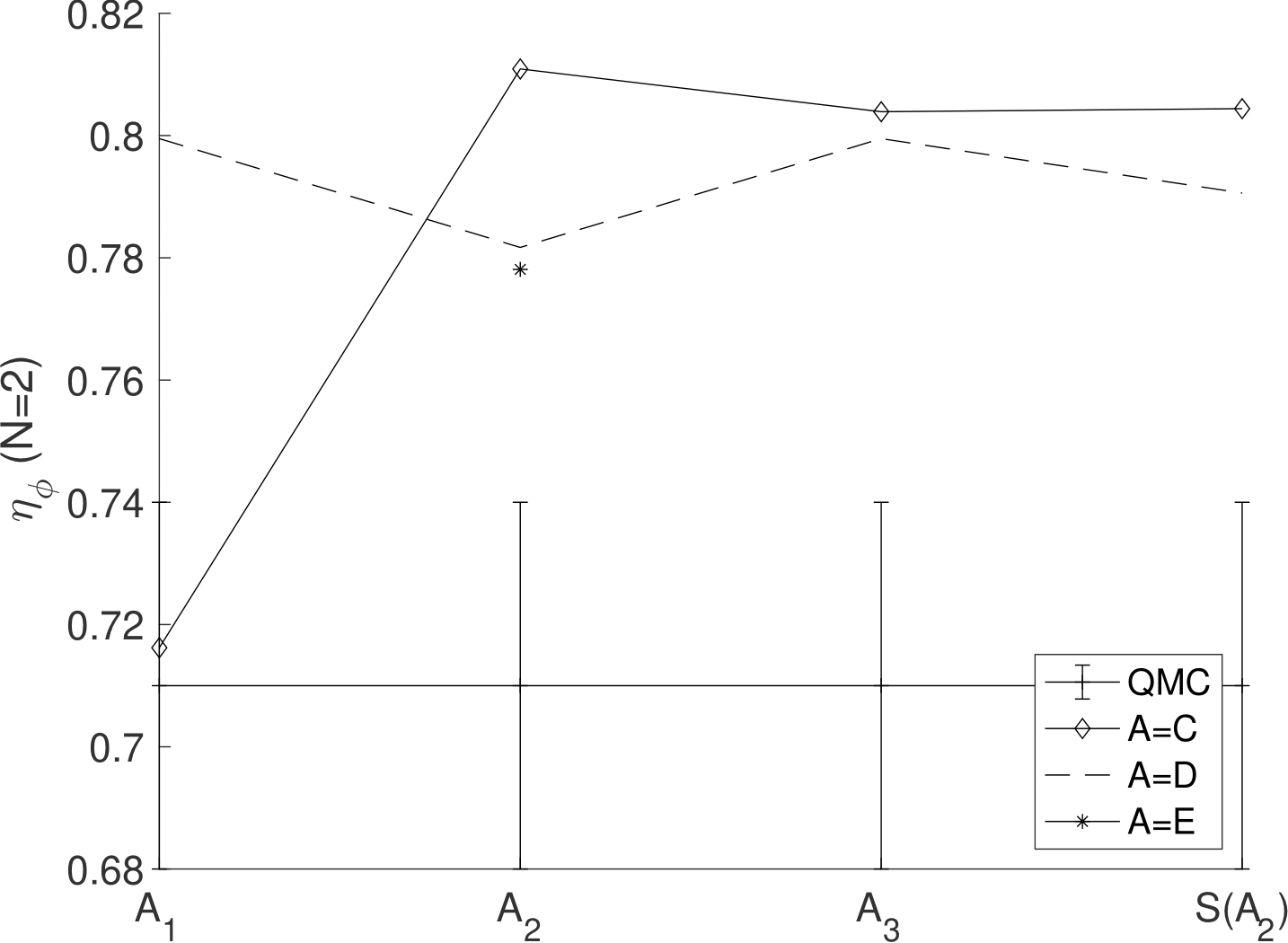}
\caption{Estimates of $\eta_{\phi}$ at successive orders.}

\end{subfigure}

\caption{Comparing Chiral XY universality class $1/\nu$ and $\eta_{\phi}$ of $N=2$  with QMC \cite{MC-XY} estimates.}
\end{figure} 
\begin{figure}[!htp]
\centering
\begin{subfigure}{0.495\textwidth}
\includegraphics[width=1\linewidth, height=6cm]{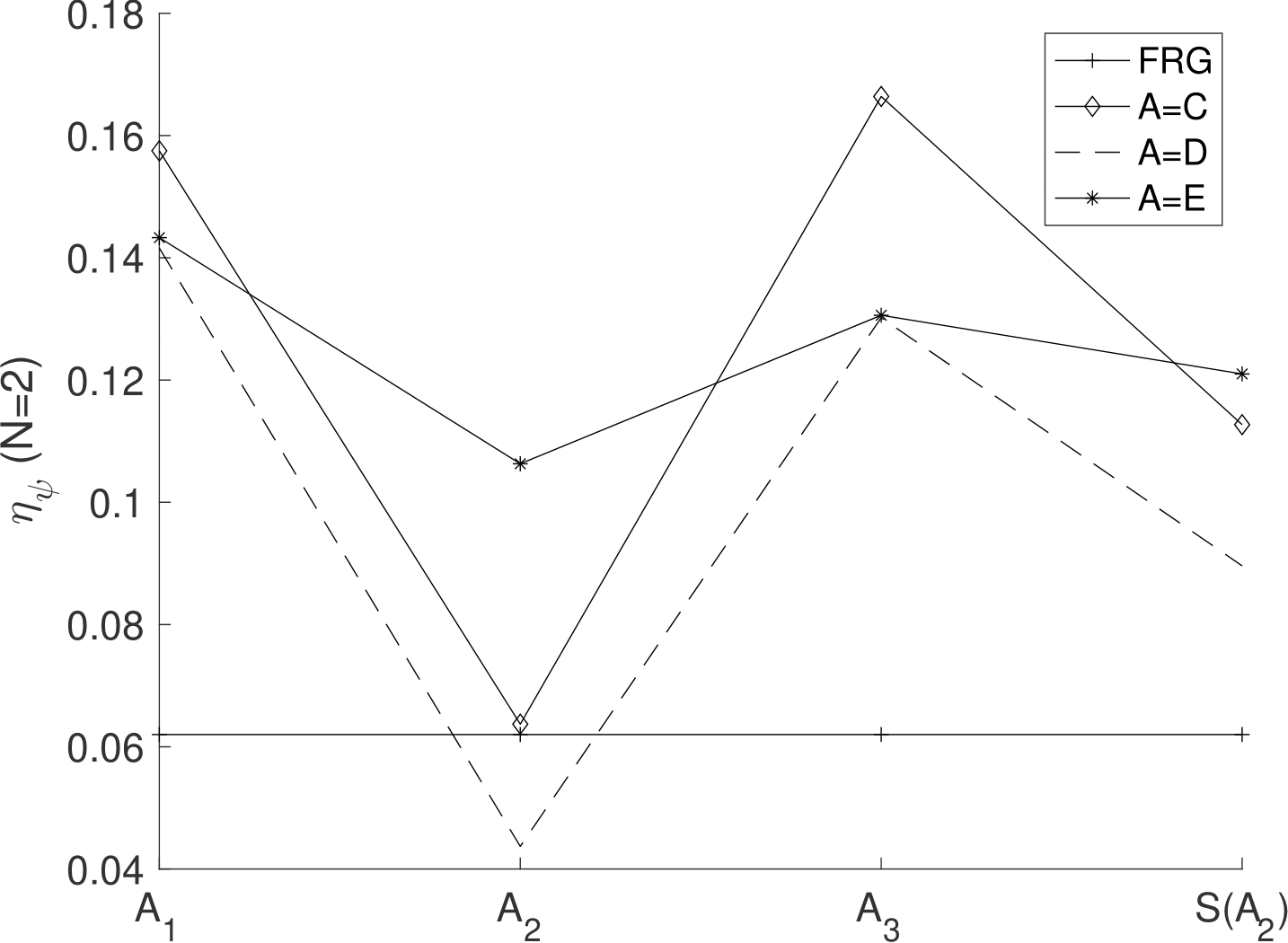} 
\caption{Estimates of $\eta_{\psi}$ at successive orders.}

\end{subfigure}
\begin{subfigure}{0.495\textwidth}
\includegraphics[width=1\linewidth, height=6cm]{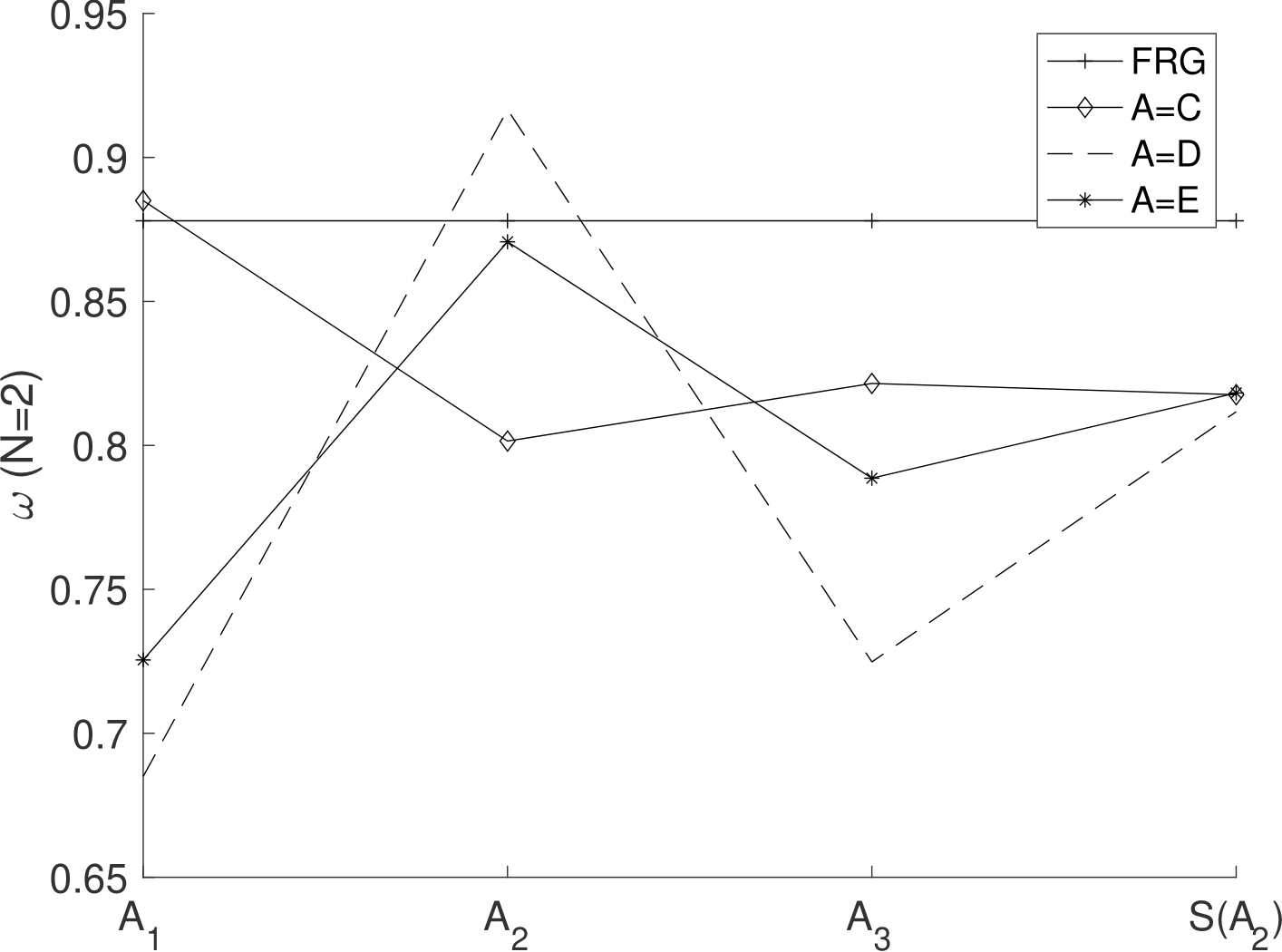}
\caption{Estimates of $\omega$ at successive orders.}

\end{subfigure}

\caption{Comparing Chiral XY universality class $\eta_{\psi}$ and $\omega$ of $N=2$  with FRG \cite{FRG-XY} estimates.}
\end{figure} We obtain the estimates of critical exponents \cite{4loopgny} \begin{subequations}
    \begin{align}
        1/\nu &=2 - \epsilon + 0.3333 \epsilon^2 - 0.8569 \epsilon^3 + 2.7629 \epsilon^4, \\
         \eta_{\phi} &= \eta_{\psi}= \epsilon/3, \\
         \omega& =\epsilon-0.3333\epsilon +0.8569\epsilon^3 -2.7629\epsilon^4,
    \end{align}
\end{subequations} for $N=1/2$ in $d=2+1$. Estimates for $1/\nu$ and $\omega$ are illustrated, compared with predictions from CB \cite{PRL-CB-XY} in Fig.s 18(a) and 18(b), respectively. \begin{figure}[!htp]
\centering
\begin{subfigure}{0.495\textwidth}
\includegraphics[width=1\linewidth, height=6cm]{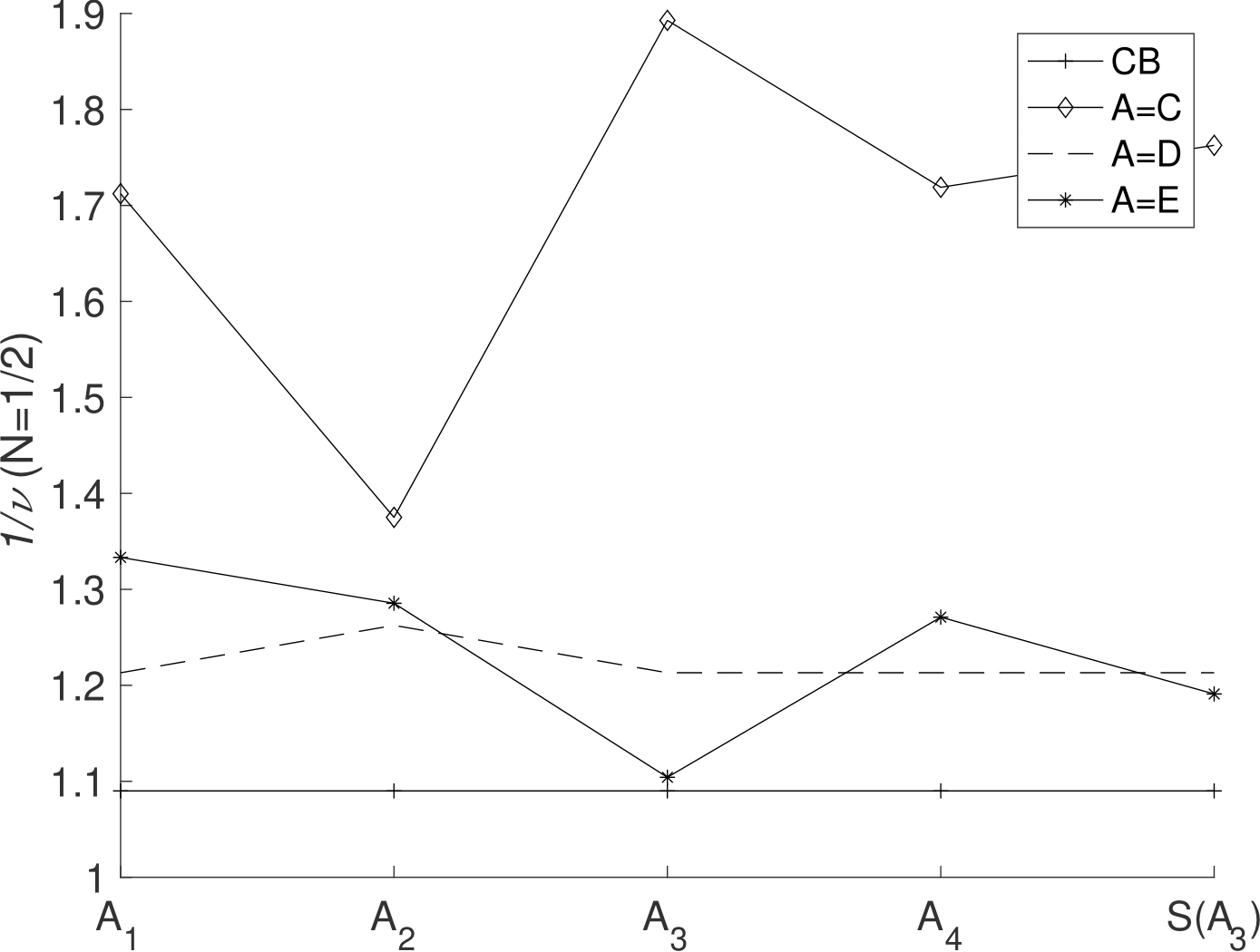} 
\caption{Estimates of $1/\nu$ at successive orders.}

\end{subfigure}
\begin{subfigure}{0.495\textwidth}
\includegraphics[width=1\linewidth, height=6cm]{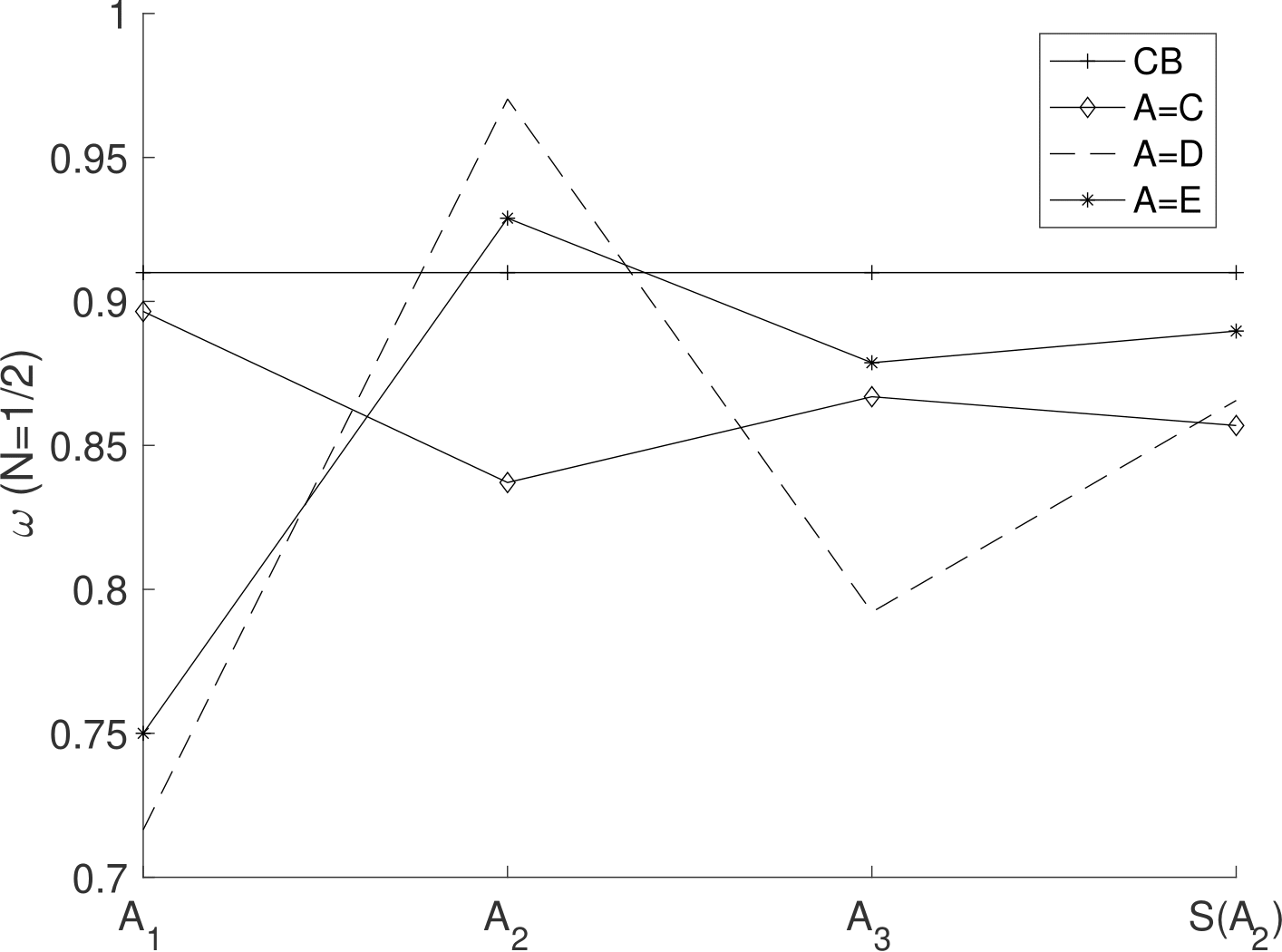}
\caption{Estimates of $\omega$ at successive orders.}

\end{subfigure}

\caption{Comparing Chiral XY universality class $1/\nu$ and $\omega$ of $N=1/2$  with CB \cite{PRL-CB-XY} estimates.}
\end{figure}  The estimated values in Table \rom{3} are comparable with predictions from other interesting field-theoretic studies of FRG \cite{FRG-XY}, QMC \cite{MC-XY}, CB \cite{PRL-CB-XY} and is compatible with Pad\'e resummation \cite{4loopgny}. 
\begingroup \scriptsize
\setlength{\tabcolsep}{.75pt} 
\renewcommand{\arraystretch}{1}
\begin{table}[htp!]
\small
\begin{center}
\caption{Critical exponents of Chiral XY universality class $1/\nu$, $\eta_{\phi}$, $\eta_{\psi}$ and $\omega$ for $N=2,1/4$ . Our values are compared with recent literature.}

\begin{tabular}{ | c | c | c | c | c | }
\hline  
$N$ & $1/\nu$ & $\eta_{\phi}$ &  $\eta_{\psi}$ & $\omega$ \\ 
\hline
 2 & \begin{tabular}{c c c c c}
      &  0.7890 ($S(C_3)$) \\
      &  0.946(26) ($S(D_3)$) \\
      & 1.0699(28) ($S(E_3)$)\\
      & 0.840, 0.841  \cite{4loopgny} \\
      & 0.862 \cite{FRG-XY} (FRG) \\
      & 1.06(5) \cite{MC-XY} (QMC)
 \end{tabular} & \begin{tabular}{c c c c c}
      & 0.8044(37) ($S(C_2)$)\\
      &  0.791(13) ($S(D_2)$) \\
      & 0.7781 ($E_2$)\\
      & 0.7079, 0.6906 \cite{4loopgny}\\
     & 0.88 \cite{FRG-XY} (FRG) \\
      & 0.71(3) \cite{MC-XY} (QMC)
 \end{tabular} & \begin{tabular}{c c c c c}
      &  0.113(78) ($S(C_2)$) \\
      &  0.121(10) ($S(D_2)$) \\
      &  0.121(17) ($S(E_2)$) \\
      & 0.117, 0.108 \cite{4loopgny} \\
      & 0.062 \cite{FRG-XY} (FRG)
 \end{tabular} & \begin{tabular}{c c c c c}   
      &  0.818(12) ($S(C_2)$) \\
      &  0.81(14) ($S(D_2)$) \\
      &  0.818(56) ($S(E_2)$) \\
      & 0.796, 0.780 \cite{4loopgny} \\
      & 0.878 \cite{FRG-XY} (FRG)
 \end{tabular} \\
 \hline 
 1/2 & \begin{tabular}{c c c c c}
      & 1.76(11) ($S(C_3)$) \\
      & 1.237(37) ($S(D_3)$)\\
      & 1.19(12) ($S(E_3)$)\\
      & 1.128,1.130 \cite{4loopgny} \\
      & 1.090 \cite{PRL-CB-XY} (CB) 
 \end{tabular} & \begin{tabular}{c c c}
      &1/3 \\
      & 1/3 \cite{4loopgny}\\
      & 1/3 \cite{PRL-CB-XY} (CB)
 \end{tabular} & \begin{tabular}{c c c}
    &1/3 \\
      & 1/3 
      \cite{4loopgny} \\
        & 1/3 \cite{PRL-CB-XY} (CB)
 \end{tabular} & \begin{tabular}{c c c}
      & 0.857(20) ($S(C_2)$) \\
      & 0.86(13) ($S(D_2)$) \\
      & 0.890(31) ($S(E_2)$)\\
      & 0.872, 0.870\cite{4loopgny}\\
      & 0.910 \cite{PRL-CB-XY} (CB)
 \end{tabular}\\
 \hline

\end{tabular}

\label{table 18}
\end{center}
\end{table} \normalsize
\subsection{Chiral Heisenberg universality class}
In Chiral Heisenberg model $SU(2)$ symmetry is broken where the description of eight component spinors ($N=2$) can correspond to transition towards an antiferromagnetic spin-density wave state in graphene and related materials \cite{prl-heis1,prl-heis2,prl-heis3}. In this case, it is interesting to note that our precise estimates of critical exponents \cite{4loopgny} \begin{subequations}
    \begin{align}
        1/\nu &=2 - 1.527 \epsilon + 0.4076 \epsilon^2 - 0.8144 \epsilon^3 + 2.001 \epsilon^4, \\
         \eta_{\phi} &= 0.8 \epsilon + 0.1593 \epsilon^2 + 0.02381 \epsilon^3 + 0.2103 \epsilon^4, \\
         \eta_{\psi} &= 0.3 \epsilon - 0.05760 \epsilon^2 - 0.1184 \epsilon^3 + 0.04388 \epsilon^4, \\
         \omega& = \epsilon - 0.4830 \epsilon^2 + 0.9863 \epsilon^3 - 2.627 \epsilon^4,
    \end{align}
\end{subequations} for $N=2$ in $d=2+1$ are more in comparison with previous predictions from FRG \cite{FRG-heis} and QMC \cite{QMC-heis41,QMC-heis62} than the simple Pad\'e estimates \cite{4loopgny} (Table \rom{4}). These estimates at consecutive orders for $1/\nu$, $\eta_{\phi}$, $\eta_{\psi}$ and $\omega$ are illustrated, compared with predictions from QMC \cite{QMC-heis62,QMC-heis41} in Fig.s 19(a), 19(b), 20(a) and 20(b), respectively. 
\begin{figure}[!htp]
\centering
\begin{subfigure}{0.495\textwidth}
\includegraphics[width=1\linewidth, height=6cm]{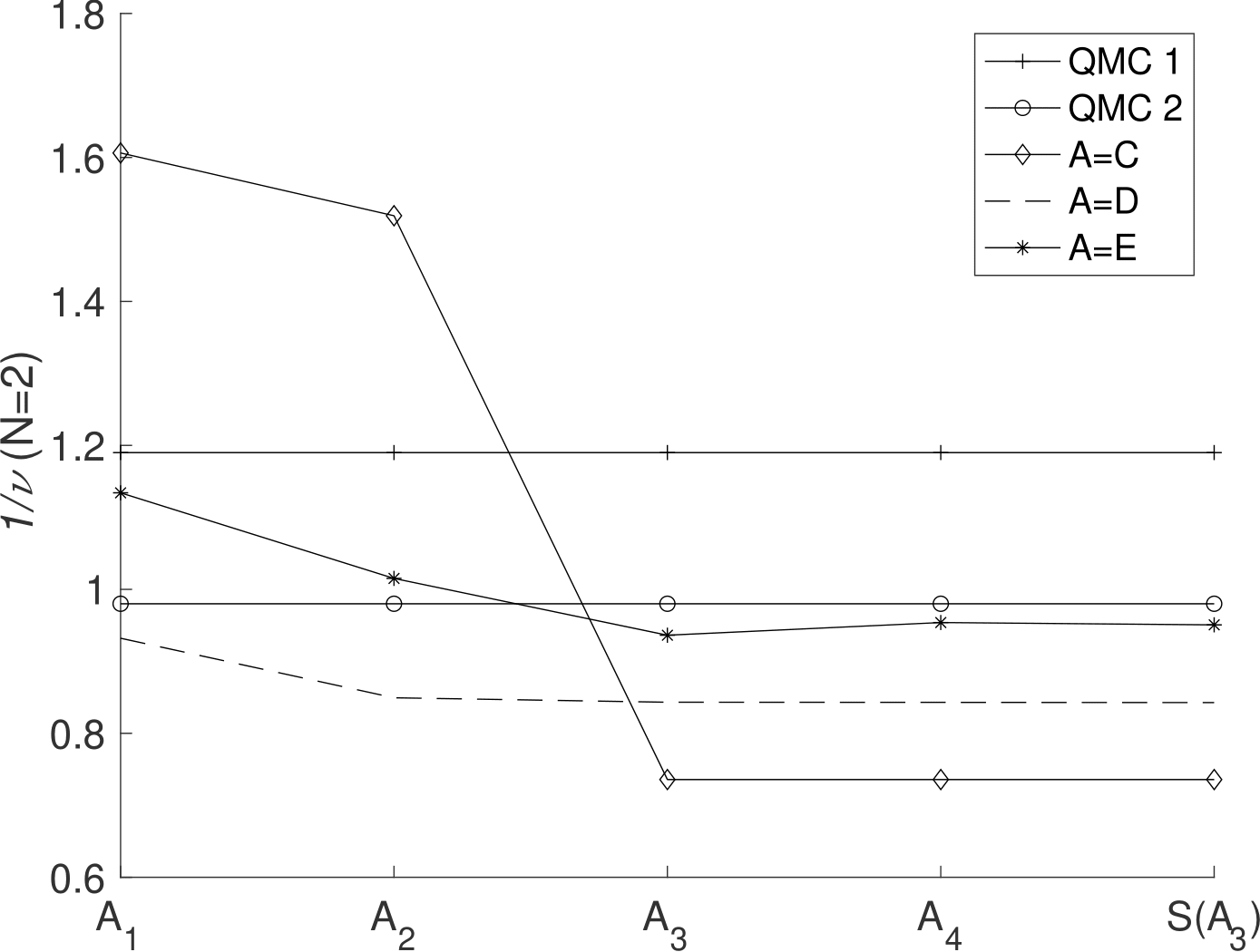} 
\caption{Estimates of $1/\nu$ at successive orders.}

\end{subfigure}
\begin{subfigure}{0.495\textwidth}
\includegraphics[width=1\linewidth, height=6cm]{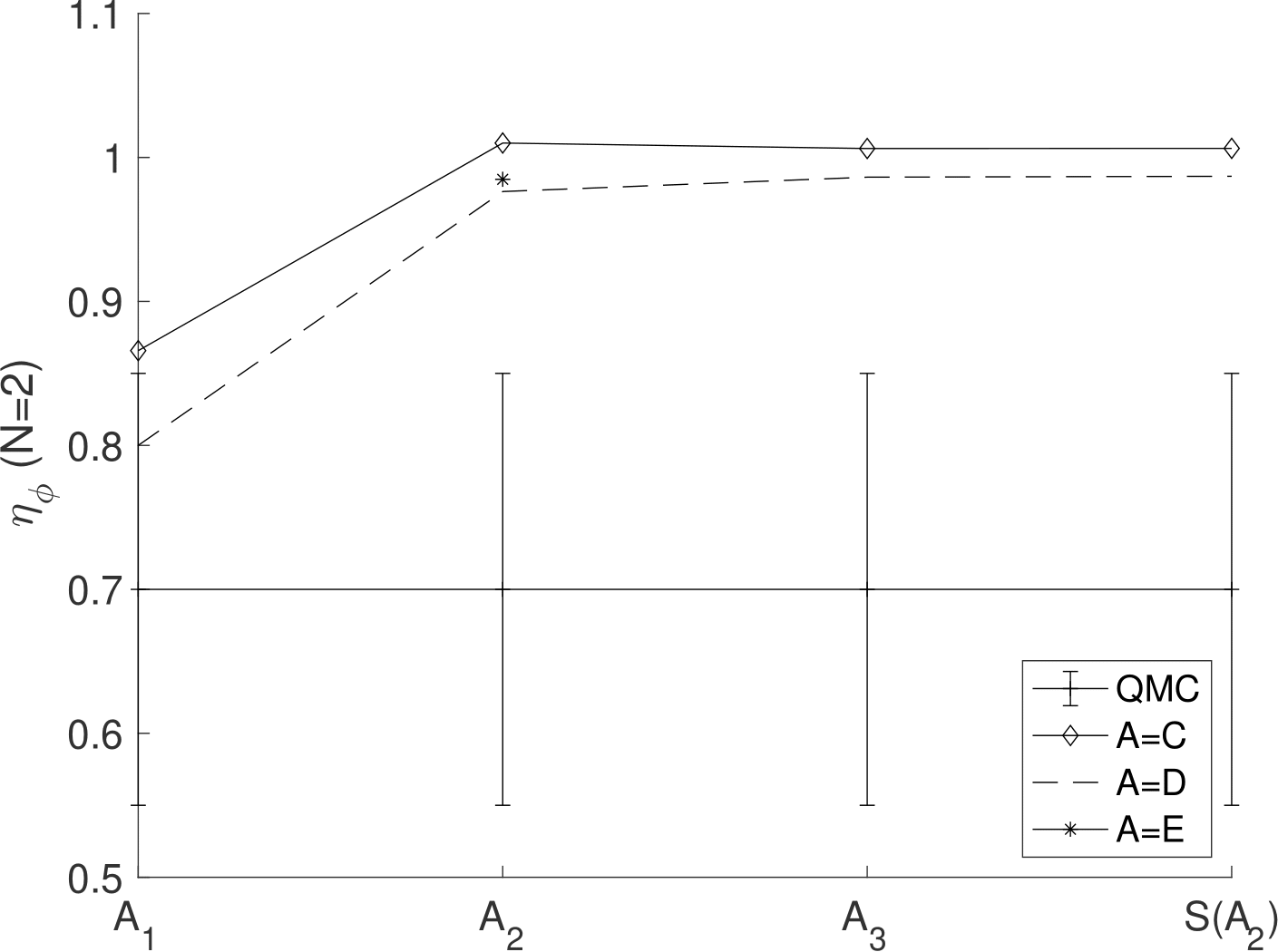}
\caption{Estimates of $\eta_{\phi}$ at successive orders.}

\end{subfigure}

\caption{Comparing Chiral Heisenberg universality class $1/\nu$ and $\eta_{\phi}$ of $N=2$  with QMC \cite{QMC-heis41,QMC-heis62} estimates.}
\end{figure} 
\begin{figure}[!htp]
\centering
\begin{subfigure}{0.495\textwidth}
\includegraphics[width=1\linewidth, height=6cm]{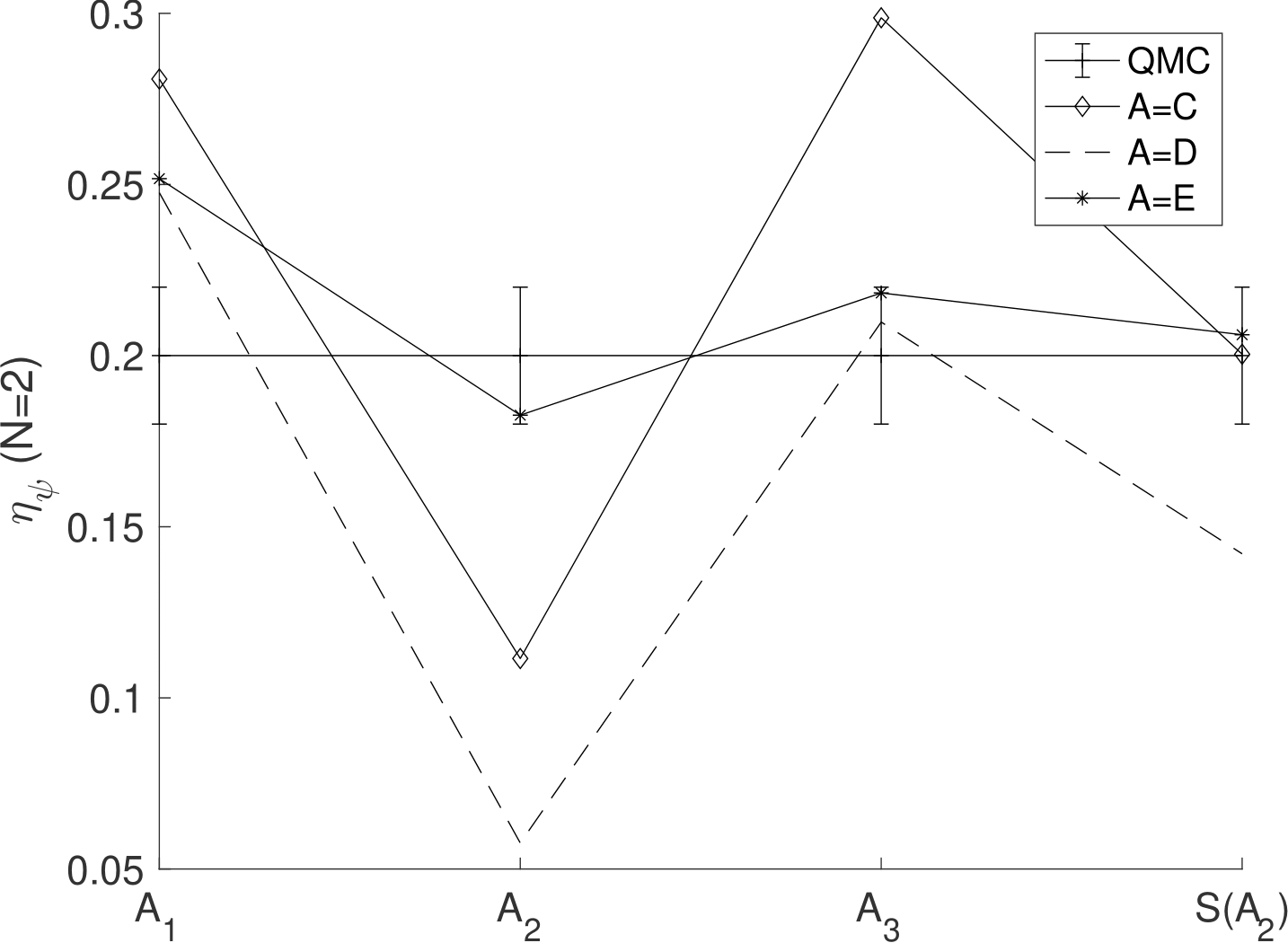} 
\caption{Estimates of $\eta_{\psi}$ at successive orders.}

\end{subfigure}
\begin{subfigure}{0.495\textwidth}
\includegraphics[width=1\linewidth, height=6cm]{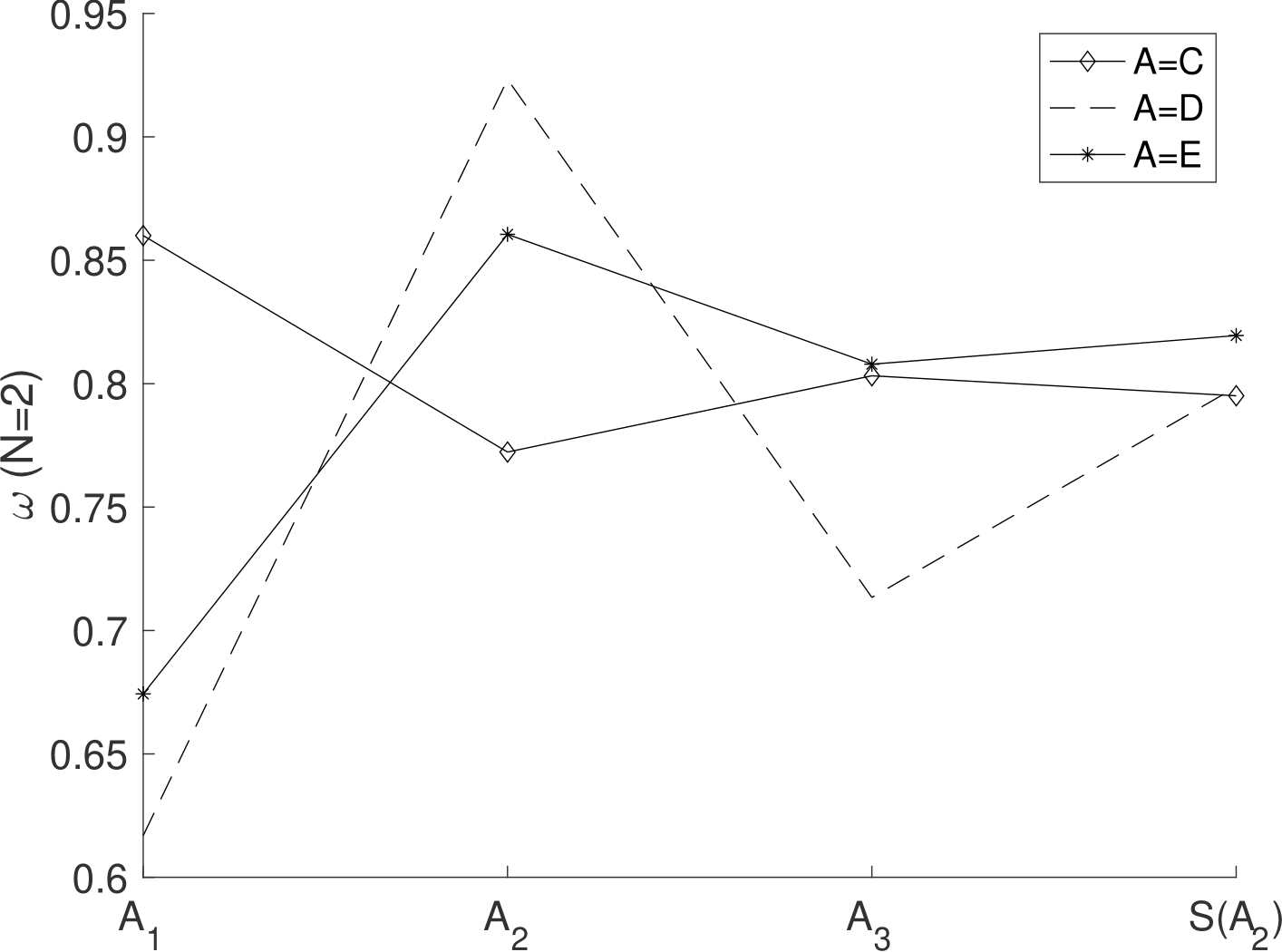}
\caption{Estimates of $\omega$ at successive orders.}

\end{subfigure}

\caption{Comparing Chiral Heisenberg universality class $\eta_{\psi}$ and $\omega$ of $N=2$  with QMC \cite{QMC-heis62} estimates.}
\end{figure}
\begingroup \scriptsize
\setlength{\tabcolsep}{.75pt} 
\renewcommand{\arraystretch}{1}
\begin{table}[htp!]
\small
\begin{center}
\caption{Critical exponents of Chiral Heisenberg universality class $1/\nu$, $\eta_{\phi}$, $\eta_{\psi}$ and $\omega$ for $N=2$ . Our values compared with recent literature.}

\begin{tabular}{ | c | c | c | c | c | }
\hline  
$N$ & $1/\nu$ & $\eta_{\phi}$ &  $\eta_{\psi}$ & $\omega$ \\ 
\hline
 2 & \begin{tabular}{c c c c c} 
      &  0.7358 ($S(C_3)$) \\
      &  0.8427(33) ($S(D_3)$) \\
      & 0.951(10) ($S(E_3)$)\\
      & 0.6426, 0.6447  \cite{4loopgny} \\
      & 0.795 \cite{FRG-heis} (FRG) \\ 
      & 0.98 \cite{QMC-heis62} (QMC) \\
      & 1.19 \cite{QMC-heis41} (QMC)
 \end{tabular} & \begin{tabular}{c c c c c}
      & 1.0063(19) ($S(C_2)$) \\
      & 0.9868(53) ($S(D_2)$)\\
      & 0.9848 ($E_2$)\\
      & 0.9985, 0.9563 \cite{4loopgny}\\
     & 1.032 \cite{FRG-heis} (FRG) \\ 
      & 0.70(15) \cite{QMC-heis41} (QMC)
 \end{tabular} & \begin{tabular}{c c c c c}
     &  0.20(14) ($S(C_2)$) \\
      &  0.14(11) ($S(D_2)$) \\
      &  0.206(24) ($S(E_2)$) \\
      & 0.1833, 0.1560 \cite{4loopgny} \\
      & 0.071 \cite{FRG-heis} (FRG) \\ 
      & 0.20(2) \cite{QMC-heis62} (QMC) 
 \end{tabular} & \begin{tabular}{c c c c c}
      & 0.795(20) ($S(C_2)$) \\
      & 0.79(15) ($S(D_2)$) \\
      & 0.820(32) ($S(E_2)$) 
 \end{tabular} \\
 \hline 
 \end{tabular}

\label{table 19}
\end{center}
\end{table} \normalsize
\section{Conclusion}
    Simple techniques were implemented on RG perturbative expansions of $O(n)$-symmetric models and Gross-Neveu-Yukawa models to better define the nature of classical and quantum phase transitions. Precise critical parameters were derived in such systems from methods using continued functions. Only the first few terms in the perturbation series are used, and methods are tried without using arbitrarily free parameters which influence the convergence. Continued exponential was implemented on perturbative low-temperature expansions and position-space renormalization scheme of the Ising model to calculate critical exponents corresponding to the system. \\
    One can further implement this convergence behaviour of continued functions on any wide range of perturbation methods to improve the convergence, especially when only a few terms are available in the divergent series. However, the accuracy of values we obtain from continued functions is only for small perturbation parameters, especially $\epsilon=1$, which makes it an ideal method to study classical systems with 3 dimensions and quantum systems with 2+1 dimensions. Further, to improve it for larger perturbation parameters, one can try to use these continued functions to interpolate from both weak and strong coupling limits using the large-order asymptotic behaviour of the perturbation coefficients, if available. The exact and unique convergent properties of an individual continued function can be further studied more rigorously based on its limits of applicability and accuracy.

\bibliographystyle{ieeetr}
\bibliography{sample.bib}

\end{document}